\documentclass{article}
\usepackage{geometry}
\usepackage{amsmath}
\usepackage{amsfonts}
\usepackage{amstext}
\usepackage{amssymb}
\usepackage{amsthm}
\usepackage{eufrak}
\newtheorem{lemma}{Lemma}

\newtheorem{theorem}{Theorem}
\newtheorem{corollary}{Corollary}
\theoremstyle{definition}
\newtheorem{definition}{Definition}

\theoremstyle{remark}
\newtheorem{remark}{Remark}

\usepackage{color}
\usepackage{graphicx} 

\begin{document}

\begin{center}
 \large \bf{Distance to the line in the Heston model}
\end{center}
\vspace{0.3cm}
\begin{center}
\bf Archil Gulisashvili \rm
\end{center}
\begin{center}
\scriptsize\it Department of Mathematics, Ohio University, Athens, OH 45701, USA \\
E-mail address: \rm gulisash@ohio.edu 
\end{center}
 \normalsize\rm
\vspace{0.1in}
\begin{center} 
This article is dedicated to the memory of Peter Laurence 
\end{center}
---------------------------------------------------------------------------------------------------------------------------------
\\
\bf Abstract  \rm 
\\
The main object of study in the paper is the distance from a point to a line in the Riemannian manifold associated with the Heston model. We reduce the problem of computing such a distance to certain minimization problems for functions of one variable over finite intervals. One of the main ideas in this paper is to use a new system of coordinates in the Heston manifold and the level sets associated with this system. In the case of a vertical line, the formulas for the distance to the line are rather simple. For slanted lines, the formulas are more complicated, and a more subtle analysis of the level sets intersecting the given line is needed.
We also find simple formulas for the Heston distance from a point to a level set. As a natural application, we use the formulas obtained in the present paper to compute the small maturity limit of the implied volatility in the 
correlated Heston model.   
\vspace{0.1in}
\vspace{0.1in}
\\
---------------------------------------------------------------------------------------------------------------------------------
\section{Introduction}
In this paper, we study a special Riemannian manifold. We call it the Heston manifold because it is intimately related to the Heston model of financial mathematics. 

The Heston model is one of the classical stock price models with stochastic volatility. The stock price process $S$ and the variance process $V$ in the Heston model satisfy the following system of stochastic differential equations:
\begin{equation}
\left\{\begin{array}{ll}
dS_t=rS_tdt+\sqrt{V_t}S_tdW_t \\
dV_t=(a-bV_t)dt+c\sqrt{V_t}dZ_t,
\end{array}
\right.
\label{E:Heston1}
\end{equation}
where $a\ge 0$, $b\ge 0$, $c> 0$, and $r\ge 0$ is the interest rate. In (\ref{E:Heston1}), $W$ and $Z$ are correlated standard Brownian motions such that 
$d\langle W,Z\rangle_t=\rho dt$ with $\rho\in(-1,1)$. 
The Heston model was introduced in \cite{H}. We refer the interested reader to \cite{FPSS,G,HL,R} for more information on the Heston model.

For the sake of simplicity, we assume throughout the paper that $r=0$. Let us denote by $X$ the log-price process in the Heston model defined by $X=\log S$. Then the model in (\ref{E:Heston1}) takes the following form: \\
$$
\left\{\begin{array}{ll}
dX_t=-\frac{1}{2}V_tdt+\sqrt{V_t}dW_t \\
dV_t=(a-bV_t)dt+c\sqrt{V_t}dZ_t.
\end{array}
\right.
$$
The state space for the process $(X,V)$ is the closed half-plane
$
{\cal H}=\left\{(x,v)\in\mathbb{R}^2:v\ge 0\right\}$.
The initial condition for the two-dimensional process $(X,V)$ will be denoted by $(x_0,v_0)$.

The Riemannian metric form associated with the uncorrelated Heston model, that is, the model with $\rho=0$, is defined on the interior ${\cal H}^{\circ}$
of the closed half-plane ${\cal H}$ as follows:
\begin{equation}
ds^2=v^{-1}\left(dx^2+dv^2\right).
\label{E:Hm}
\end{equation} 
This form generates the Riemannian distance $d_H$ on ${\cal H}$. We call the open half-plane
${\cal H}^{\circ}$, equipped with the metric form defined above, the Heston Riemannian manifold (see
\cite{GL} for more details). The line $\{(x,v):v=0\}$ is the boundary of the Heston manifold, and the manifold is incomplete.
\begin{remark}\label{R:HD} \rm
Riemannian metrics similar to that in (\ref{E:Hm}) also appear in other fields of mathematics. For example, P. Daskalopoulos and R. Hamilton used the Riemannian metric in the right half-plane, defined by
\begin{equation}
ds^2=(2x)^{-1}\left(dx^2+dy^2\right),
\label{E:M}
\end{equation}
to study the regularity of the interface of the evolution $p$-Laplacian equation
(see \cite{DH2}) and the porous medium equation (see \cite{DH1}). Daskalopoulos and Hamilton call the metric in (\ref{E:M}) the cycloidal metric, since all the geodesics of this metric can be obtained from the standard cycloid curve by translation and dilation, or are horizontal lines (see Proposition I.2.1 in \cite{DH1}).
\end{remark}

Methods of mathematical analysis and differential geometry found numerous applications in quantitative finance. A good source of information about such applications is the book \cite{HL} by P. Henry-Labord\`{e}re. This book also discusses the geometry of the Heston model. In \cite{GL}, the author and P. Laurence found explicit formulas for the Heston Riemannian distance between two points. The main emphasis in the present paper is on the distance from a point to a line in the Heston manifold. A new method of studying the Heston manifold is suggested in the paper. The main idea behind this method is to use the level sets associated with a new curvilinear system of coordinates in the Heston manifold (see Subsection \ref{SS:news}). We link the problem of computing the distance to a line with certain minimization problems for functions of one variable over finite intervals. A natural application of any description of the distance to the line in the Heston manifold is to the study of the small-time behavior of the implied volatility, since it is known how the leading term in the small-time asymptotic expansion of the implied volatility in the correlated Heston model is related to the distance to the line in the Heston manifold (see Section \ref{S:appii}).

In the case where $\rho\neq 0$, the Heston Riemannian distance will be denoted by $d^{(\rho)}_H$. The following equality holds for the Riemannian distance $d^{(\rho)}_H$ and the distance $d_H$ in the corresponding uncorrelated Heston model (see (7) in \cite{GL}): 
\begin{align}
&d^{(\rho)}_H((x_0,v_0),(x_1,v_1))
=\frac{1}{c}d_H\left(\left(\frac{cx_0-\rho v_0}{\sqrt{1-\rho^2}},v_0\right),\left(\frac{cx_1-\rho v_1}{\sqrt{1-\rho^2}},v_1\right)\right).
\label{E:sveli}
\end{align}

Fix real numbers $\gamma$ and $\beta$, and denote by $L_{\beta,\gamma}$ the line in the upper half-plane ${\cal H}$ given by 
$$
\left\{(x,v)\in{\cal H}:x=\beta+\gamma v,\,\,v\ge 0\right\}.
$$ 
The symbol $\widehat{D}_{\beta,\gamma}$ will stand for the distance from the point $(0,1)\in{\cal H}$ to the 
line $L_{\beta,\gamma}$ in the uncorrelated Heston model ($\rho=0$). We have
\begin{equation}
\widehat{D}_{\beta,\gamma}=\inf_{v\ge 0}\left\{d_H((0,1),(\beta+\gamma v,v))\right\}.
\label{E:H2}
\end{equation}
\begin{remark}\label{R:export} \rm
Note that the minimum in (\ref{E:H2}) can not be attained at infinity. This assertion can be obtained, using
the following two-sided estimates for the Heston distance established in 
\cite{K}, Proposition 4.3.2:
\begin{equation}
T(x_0,v_0,x_1,v_1)\le d_H((x_0,v_0),(x_1,v_1))\le 12T(x_0,v_0,x_1,v_1)
\label{E:twos1}
\end{equation}
for all $(x_0,v_0)\in{\cal H}$ and $(x_1,v_1)\in{\cal H}$,
where
\begin{equation}
T(x_0,v_0,x_1,v_1)=\frac{\sqrt{(x_0-x_1)^2+(v_0-v_1)^2}}{\sqrt{v_0}+\sqrt{v_1}
+[(x_0-x_1)^2+(v_0-v_1)^2]^{\frac{1}{4}}}.
\label{E:twos3}
\end{equation}
Indeed, (\ref{E:twos1}) shows that for all $\beta$ and $\gamma$,
$$
\lim_{v_1\rightarrow\infty}d_H((0,1),(\beta+\gamma v_1,v_1))=\infty.
$$
\end{remark}
\begin{remark}\label{R:exxo} \rm Note that (\ref{E:twos3}) implies the following equality:
\begin{equation}
T(0,1,x,v)=\frac{\sqrt{x^2+(v-1)^2}}{1+\sqrt{v}+[x^2+(v-1)^2]^{\frac{1}{4}}}
\label{E:tti}
\end{equation}
for all $x\in\mathbb{R}$ and $v\ge 0$. 
\end{remark}
\begin{remark}\label{R:equm} \rm A two-sided estimate, equivalent to that in (\ref{E:twos1}), but with no information on the constants, is formulated in part 1 of \cite{DH1} and part 2 of \cite{DH2}.
\end{remark}

Let $(x_0,v_0)\in{\cal H}$ and $(x_1,v_1)\in{\cal H}$ be points in the Heston manifold such that at least one of
them is not on the boundary. The following explicit formula for the Riemannian distance $d_H$ between those points
was obtained in \cite{GL}:
\begin{equation}
d_H\left((x_0,v_0),(x_1,v_1)\right)=
\frac{\delta}{\sin\left(\frac{\delta}{2}\right)}
\sqrt{v_1+v_0-2\sqrt{v_1v_0}\cos\left(\frac{\delta}{2}\right)},
\label{E:nes5}
\end{equation}
where $\delta=\delta((x_0,v_0),(x_1,v_1))$
is the unique solution to the equation
\begin{align}
&\frac{\left(v_1+v_0\right)
\left(\delta-\sin(\delta)\right)
-2\sqrt{v_1v_0}\left(\delta\cos\left(\frac{\delta}{2}\right)
-2\sin\left(\frac{\delta}{2}\right)\right)}
{2\sin^2\left(\frac{\delta}{2}\right)}=x_1-x_0,
\label{E:nes6}
\end{align}
satisfying the condition $-2\pi<\delta< 2\pi$.

It follows from (\ref{E:H2}), (\ref{E:nes5}), and (\ref{E:nes6}) that
\begin{align}
&\widehat{D}_{\beta,\gamma}=\inf_{v\ge 0}\left\{\frac{\delta}{\sin\left(\frac{\delta}{2}\right)}
\sqrt{v+1-2\sqrt{v}\cos\left(\frac{\delta}{2}\right)}\right\},
\label{E:ali}
\end{align}
where $\delta$ with $-2\pi<\delta< 2\pi$ is the unique solution to the following equation:
\begin{align}
&\frac{\left(v+1\right)
\left(\delta-\sin(\delta)\right)
-2\sqrt{v}\left(\delta\cos\left(\frac{\delta}{2}\right)
-2\sin\left(\frac{\delta}{2}\right)\right)}
{2\sin^2\left(\frac{\delta}{2}\right)}=\beta+\gamma v.
\label{E:nestor}
\end{align}
However, formula (\ref{E:ali}) is not very efficient numerically, since in order to use it in computations, we need to solve equation (\ref{E:nestor}) for every $v\ge 0$.

Our main objective in the present paper is to find simple and efficient formulas for the distance $\widehat{D}_{\beta,\gamma}$ to the line $L_{\beta,\gamma}$ in the uncorrelated Heston model. For vertical lines,
the main results concerning the distance to the line problem are contained in Theorem \ref{T:kp}. For right slanted 
lines, the main distance formulas can be found in Theorems \ref{T:the1} and \ref{T:di}, and in Section \ref{SS:usef}. Finally, for left slanted lines, the main results are contained in Theorem \ref{T:gammane}. 

Set
\begin{equation}
D_{\beta,\gamma}=\inf\left\{\frac{d_H((0,1),(x,v))^2}{2}:(x,v)\in L_{\beta,\gamma}\right\}.
\label{E:distance}
\end{equation}
We will see below that it is easier to work with the quantity $D_{\beta,\gamma}$. Note that 
\begin{equation}
\widehat{D}_{\beta,\gamma}=\sqrt{2D_{\beta,\gamma}},
\label{E:Synge}
\end{equation}
for all $\beta$ and $\gamma$.

The distance $d_H$ satisfies the following conditions:
\begin{equation}
d_H((x_0,v_0),(x_1,v_1))=d_H((0,v_0),(x_1-x_0,v_1))
\label{E:form2}
\end{equation}
and
\begin{equation}
d_H((\alpha x_0,\alpha v_0),(\alpha x_1,\alpha v_1))=\sqrt{\alpha}
d_H((x_0,v_0),(x_1,v_1))
\label{E:form3}
\end{equation}
for all $\alpha> 0$ (see \cite{GL}, formulas (8) and (9)). Next, using (\ref{E:sveli}),
(\ref{E:form2}), and (\ref{E:form3}) we see that
$$
d^{(\rho)}_H((x_0,v_0),(x_1,v_1))
=\frac{\sqrt{v_0}}{c}d_H\left((0,1),
\left(\frac{c(x_1-x_0)-\rho(v_1-v_0)}{v_0\sqrt{1-\rho^2}},\frac{v_1}{v_0}\right)\right).
$$
It follows from the previous formula that
\begin{equation}
d^{(\rho)}\left((x_0,v_0),L_{\beta,\gamma}\right)=\frac{\sqrt{v_0}}{c}d_H\left((0,1),L_{\xi,\eta}\right),
\label{E:sova}
\end{equation}
where
\begin{equation}
\xi=\frac{c\beta-cx_0+\rho v_0}{v_0\sqrt{1-\rho^2}}\quad\mbox{and}\quad\eta=\frac{c\gamma-\rho}{\sqrt{1-\rho^2}}.
\label{E:filin}
\end{equation}
\begin{remark}\label{R:sviaz} \rm Formula (\ref{E:sova}) shows that for the correlated Heston model, the problem of computing the distance from a general point $(x_0,v_0)\in{\cal H}$ to a line $L_{\beta,\gamma}$ can be reduced to a similar problem for the uncorrelated model, the point $(0,1)$, and the line $L_{\xi,\eta}$ with
$\xi$ and $\eta$ given by (\ref{E:filin}).
\end{remark}

We will next briefly overview the results obtained in the present paper. In Section \ref{S:formmain},
we compute the Heston distance from the point $(0,1)$ to a vertical line. The main result of this section is Theorem 
\ref{T:kp} that provides formulas for the distance to the vertical line in terms of minimization problems for an explicit function of one variable over a finite interval. Subsection \ref{SS:levels} is devoted to the study of the level sets of the function $\delta$. The idea of using the level sets of $\delta$ is one of the main novelties of the present paper. In Subsection \ref{SS:news}, we introduce a new system of coordinates in the Heston manifold, using the level sets of $\delta$, while in Subsection \ref{SS:levelset}, we compute the Heston distance from the point $(0,1)$ to a level set of $\delta$. The formulas for this distance, provided in Theorem \ref{T:ditol}, are especially simple. Subsection \ref{SS:dhl} deals with the distance to a horizontal line. The proof of Theorem \ref{T:kp} is completed in Subsection \ref{SS:dttl}. The problem of computing the distance from the point $(0,1)$ to a slanted line is addressed in Section \ref{S:skew}. The main results of Section \ref{S:skew} for right slanted lines are gathered in Theorem 
\ref{T:the1} formulated in Subsection \ref{SS:mainre}. The family of level sets of $\delta$, which intersect the given slanted line is described in Subsection \ref{SS:sc}. Here we also complete the proof of Theorem \ref{T:the1}. In Subsection \ref{SS:ls}, we formulate and prove Theorem \ref{T:gammane}, which provides formulas for the distance to left slanted lines. Section \ref{SS:usef} contains improvements and simplifications of the distance formulas provided in Theorem \ref{T:the1} under additional restrictions on the parameters. Section \ref{S:dsl} discusses the distance formulas for certain special lines in the Heston manifold. These lines are tangent lines to the level sets of $\delta$ at the critical points. It is interesting that the distance formulas for such lines are extremely simple. Finally, in Section \ref{S:appii}, we explain how the distance formulas are related to the small-time limit of the implied volatility in the correlated Heston model.
\section{Distance to a vertical line in the Heston manifold}\label{S:formmain}
Our main goal in the present section is to find a formula characterizing the quantity $D_{\beta,0}$. Then, using (\ref{E:Synge}),
we can compute the distance $\widehat{D}_{\beta,0}$ from the point $(0,1)$ to the vertical line $x=\beta$ in the Heston manifold. The following function will be used in the sequel:
\begin{align}
&\Lambda(x,\theta)=\frac{\theta^2}{(\theta-\sin\theta)^2}[(\theta-\sin\theta)x+2(1-\cos\theta)-\theta\sin\theta \nonumber \\
&\quad-(1-\cos\theta)\sqrt{2(\theta-\sin\theta)x+2(1-\cos\theta)-\theta^2}].
\label{E:recs}
\end{align}
The function $\Lambda$ defined in (\ref{E:recs}) is finite and real if $\theta\neq 0$ and 
$2(\theta-\sin\theta)x+2(1-\cos\theta)-\theta^2\ge 0$. We choose the positive value of the square root function in 
(\ref{E:recs}).

The next assertion is one of the main results obtained in the present paper.
It states that the number $D_{\beta,0}$ is the solution to a minimization problem for the function $\Lambda$ on an
explicitly defined finite closed interval. 
\begin{theorem}\label{T:kp}
If $0<\beta<\frac{\pi}{2}$, then
\begin{equation}
D_{\beta,0}=\min_{\{\theta:\frac{1}{11}\beta\le\theta\le 2\beta\}}
\left\{\Lambda(\beta,\theta)\right\},
\label{E:whos1}
\end{equation}
while if $\frac{\pi}{2}\le\beta<\infty$, then
\begin{equation}
D_{\beta,0}=\min_{\{\theta:\frac{1}{7}\le\theta\le\pi\}}
\left\{\Lambda(\beta,\theta)\right\}.
\label{E:whos2}
\end{equation}
\end{theorem}

The intervals in (\ref{E:whos1}) and (\ref{E:whos2}) contain unique critical points associated with the minimization problems described in the formulation of Theorem \ref{T:kp}. Note that the end points of the interval appearing in formula (\ref{E:whos2}) do not depend on the parameter $\beta$. The proof of Theorem \ref{T:kp} will be given in the next subsections. We first develop the necessary machinery that will be used in the proof. The level sets of the function $(x,v)\mapsto\delta((0,1),(x,v))$ play an important role in this section and in the restg of the paper. 
The proof of Theorem \ref{T:kp} will be completed at the very end of Subsection \ref{SS:dttl}. Shorter intervals in the minimization problems in (\ref{E:whos1}) and (\ref{E:whos2}) can also be found (see Corollaries \ref{C:reduction1} and \ref{C:reduction2}). However, the end points of the intervals in those corollaries are more complicated.
\subsection{Useful formulas}\label{SS:delta}
It follows from (\ref{E:nes5}) and (\ref{E:nes6}) that
\begin{equation}
\frac{d_H((0,1),(x,v))^2}{2}=\frac{\delta^2}{2\sin^2\frac{\delta}{2}}
\left(v+1-2\sqrt{v}\cos\frac{\delta}{2}\right),
\label{E:synge}
\end{equation}
where $\delta=\delta(x,v)$ is the unique solution with $-2\pi<\delta<2\pi$ of the following transcendental equation:
$f(v,\delta)=x$, where
\begin{equation}
f(v,\delta)=\frac{(v+1)(\delta-\sin\delta)+2\sqrt{v}(2\sin\frac{\delta}{2}-\delta\cos\frac{\delta}{2})}
{2\sin^2\frac{\delta}{2}}.
\label{E:forma1}
\end{equation}
In (\ref{E:forma1}), the positive value of $\sqrt{v}$ is used. The function $f$ in (\ref{E:forma1}) is strictly increasing and convex on the interval $[0,2\pi)$ (see \cite{GL}).

The next lemma provides equivalent formulas for the function on the left-hand side of (\ref{E:synge}).
\begin{lemma}\label{L:ssy}
The following formulas are valid:
\begin{equation}
\frac{d_H((0,1),(x,v))^2}{2}=\frac{\delta^2}{2\sin^2\frac{\delta}{2}}
\left((\sqrt{v}-1)^2+4\sqrt{v}\sin^2\frac{\delta}{4}\right)
\label{E:usef2}
\end{equation}
and
\begin{equation}
\frac{d_H((0,1),(x,v))^2}{2}=\frac{\delta^2}{\delta-\sin\delta}\left(x
-2\sqrt{v}\sin\frac{\delta}{2}\right).
\label{E:simple1}
\end{equation}
\end{lemma}

\it Proof. \rm Using (\ref{E:synge}), we obtain
\begin{align*}
\frac{d_H((0,1),(x,v))^2}{2}&=\frac{\delta^2}{2\sin^2\frac{\delta}{2}}
\left(v+1-2\sqrt{v}\left[1-2\sin^2\frac{\delta}{4}\right]\right) \\
&=\frac{\delta^2}{2\sin^2\frac{\delta}{2}}
\left((\sqrt{v}-1)^2+4\sqrt{v}\sin^2\frac{\delta}{4}\right).
\end{align*}
This establishes equality (\ref{E:usef2}).

To prove (\ref{E:simple1}), we first rewrite the equation $f(v,\delta)=x$ in the following form:
$$
\frac{v+1}{2\sin^2\frac{\delta}{2}}=\frac{x\sin^2\frac{\delta}{2}-\sqrt{v}\left(2\sin\frac{\delta}{2}
-\delta\cos\frac{\delta}{2}\right)}{\sin^2\frac{\delta}{2}(\delta-\sin\delta)}.
$$
Next, using the previous formula and (\ref{E:synge}), we obtain
\begin{align*}
&\frac{d_H((0,1),(x,v))^2}{2}=\delta^2\left[\frac{x\sin^2\frac{\delta}{2}
-\sqrt{v}\left(2\sin\frac{\delta}{2}
-\delta\cos\frac{\delta}{2}\right)}{\sin^2\frac{\delta}{2}(\delta-\sin\delta)}
-\frac{\sqrt{v}\cos\frac{\delta}{2}}{\sin^2\frac{\delta}{2}}\right] \\
&=\delta^2\frac{x\sin^2\frac{\delta}{2}
-\sqrt{v}\left(2\sin\frac{\delta}{2}
-\delta\cos\frac{\delta}{2}+\cos\frac{\delta}{2}(\delta-\sin\delta)\right)}{\sin^2\frac{\delta}{2}(\delta-\sin\delta)} =\delta^2\frac{x\sin^2\frac{\delta}{2}
-2\sqrt{v}\sin^3\frac{\delta}{2}}{\sin^2\frac{\delta}{2}(\delta-\sin\delta)}.
\end{align*}

Now, it is clear that (\ref{E:simple1}) holds.
\begin{lemma}\label{L:oeq}
For all $v\ge 0$ and $-2\pi<\delta<2\pi$,
\begin{equation}
f(v,\delta)=\frac{(\sqrt{v}-1)^2(\delta-\sin\delta)+4\sqrt{v}\sin^2\frac{\delta}{4}
\left(\delta+2\sin\frac{\delta}{2}\right)}{2\sin^2\frac{\delta}{2}}.
\label{E:thef}
\end{equation}
\end{lemma}

\it Proof. \rm We have 
\begin{align}
&(v+1)(\delta-\sin\delta)+2\sqrt{v}\left(2\sin\frac{\delta}{2}-\delta\cos\frac{\delta}{2}\right) \nonumber \\
&=(\sqrt{v}-1)^2(\delta-\sin\delta)+2\sqrt{v}\left(\delta-\sin\delta+2\sin\frac{\delta}{2}
-\delta\cos\frac{\delta}{2}\right) \nonumber \\
&=(\sqrt{v}-1)^2(\delta-\sin\delta)+2\sqrt{v}\left(\delta\left(1-\cos\frac{\delta}{2}\right)+2\sin\frac{\delta}{2}
\left(1-\cos\frac{\delta}{2}\right)\right) \nonumber \\
&=(\sqrt{v}-1)^2(\delta-\sin\delta)+4\sqrt{v}\sin^2\frac{\delta}{4}\left(\delta+2\sin\frac{\delta}{2}\right).
\label{E:nf}
\end{align}

Finally, using (\ref{E:forma1}) and (\ref{E:nf}), we obtain (\ref{E:thef}).

In the next lemma, we find an invertible majorant for the function $f$. The proof uses formula (\ref{E:thef}).
\begin{lemma}\label{L:major}
The following inequality holds for all $v\ge 0$ and $0<\delta< 2\pi$:
$f(v,\delta)\le g(v,\delta)$, where the function $g$ is defined as follows:
\begin{equation}
g(v,\delta)=\left\{\begin{array}{ll}
\frac{\pi^2}{12}(v+\sqrt{v}+1)\delta,\quad\mbox{if}\quad 0<\delta\le\pi \\
\frac{\pi^8}{12}(v+\sqrt{v}+1)(2\pi-\delta)^{-5},\quad\mbox{if}\quad \pi\le\delta< 2\pi.
\end{array}
\right.
\label{E:kabel}
\end{equation}
\end{lemma}

\it Proof. \rm We have 
\begin{equation}
\sin\delta\le\delta,\quad 0<\delta< 2\pi,
\label{E:trig0}
\end{equation}
\begin{equation}
\delta-\sin\delta\le\frac{1}{6}\delta^3,\quad 0<\delta< 2\pi,
\label{E:trig1}
\end{equation}
\begin{equation}
\sin\frac{\delta}{2}\ge\frac{\delta}{\pi},\quad 0<\delta\le\pi,
\label{E:trig2}
\end{equation}
and
\begin{equation}
\sin\frac{\delta}{2}\ge 2-\frac{\delta}{\pi},\quad \pi\le\delta< 2\pi.
\label{E:trig3}
\end{equation}

Let $0<\delta\le\pi$. Then, using (\ref{E:thef}), (\ref{E:trig0}), (\ref{E:trig1}), and (\ref{E:trig2}), we obtain
\begin{align*}
f(v,\delta)&\le\frac{\pi^2\left[6^{-1}(\sqrt{v}-1)^2\delta^3+2^{-1}\sqrt{v}\delta^3\right]}{2\delta^2} \\
&=\frac{\pi^2}{12}(v+\sqrt{v}+1)\delta=g(v,\delta).
\end{align*}
Now, let $\pi\le\delta< 2\pi$. Then (\ref{E:thef}), (\ref{E:trig0}), (\ref{E:trig1}), and (\ref{E:trig3}) imply
\begin{align}
f(v,\delta)&\le\frac{\pi^2\left[6^{-1}(\sqrt{v}-1)^2\delta^3+2^{-1}\sqrt{v}\delta^3\right]}
{2(2\pi-\delta)^2} \nonumber \\
&=\frac{\pi^2}{12}(v+\sqrt{v}+1)\frac{\delta^3}{(2\pi-\delta)^2}.
\label{E:trig5}
\end{align}

Our next goal is to show that
\begin{equation}
\frac{\delta^3}{(2\pi-\delta)^2}\le\frac{\pi^6}{(2\pi-u)^5},
\label{E:trigg1}
\end{equation}
for all $\pi\le\delta< 2\pi$. It is easy to see that it suffices to prove the following inequality:
\begin{equation}
\frac{u^3}{(2-u)^2}\le\frac{1}{(2-u)^5},
\label{E:trigg2}
\end{equation}
for all $1\le u< 2$. The inequality in (\ref{E:trigg2}) is equivalent to the inequality
$u(2-u)\le 1$, $1\le u< 2$, which is clearly true. This establishes (\ref{E:trigg1}). Moreover, it follows from
(\ref{E:trig5}) and (\ref{E:trigg1}) that
\begin{equation}
f(v,\delta)\le\frac{\pi^8}{12}(v+\sqrt{v}+1)(2\pi-\delta)^{-5}=g(v,\delta).
\label{E:trigg3}
\end{equation}
Now, Lemma \ref{L:major} follows from (\ref{E:trigg1}) and (\ref{E:trigg3}).

The next statement provides a useful estimate from below for the parameter $\delta$ in the Heston model.
\begin{lemma}\label{L:useful}
Let $x> 0$ and $v\ge 0$. Then $\delta((0,1),(x,v))\ge h(x,v)$, where
\begin{align*}
&h(x,v)=\left\{\begin{array}{ll}
\frac{12 x}{\pi^2(v+\sqrt{v}+1)},\,\,\mbox{if}\,\,0< x\le\frac{\pi^3}{12}(v+\sqrt{v}+1) \\
2\pi-\left[\frac{\pi^8(v+\sqrt{v}+1)}{12x}\right]^{\frac{1}{5}},\,\,\mbox{if}\,\, \frac{\pi^3}{12}(v+\sqrt{v}+1)\le x<\infty.
\end{array}
\right.
\end{align*}
\end{lemma}

\it Proof. \rm It is easy to see that the function $\delta\mapsto g(v,\delta)$, where $g$ is defined 
by (\ref{E:kabel}), is strictly increasing and continuous on the interval $(0,2\pi)$. It is also clear that this function maps $(0,2\pi)$ onto $(0,\infty)$. Since $f(v,\theta)=x$,
we have $\theta=f^{-1}(v,\cdot)(x)$. Moreover, the estimate 
$f(v,\delta)\le g(v,\delta)$
(see Lemma \ref{L:major}) implies that
$g^{-1}(v,\cdot)(x)\le f^{-1}(v,\cdot)(x)$.
Therefore, $\theta\ge g^{-1}(v,\cdot)(x)$, and the 
inequality in Lemma \ref{L:useful} with $h=g^{-1}(v,\cdot)$ follows from the previous estimate and (\ref{E:kabel}).

This completes the proof of Lemma \ref{L:useful}.
\subsection{Level sets of the function $\delta$}\label{SS:levels}
The level sets $\Gamma_{\theta}$, $-2\pi<\theta<2\pi$, of the function $(x,v)\mapsto\delta((0,1),(x,v))$ play an important role in the present paper.
The definition of the set $\Gamma_{\theta}$ is as follows:
$\Gamma_{\theta}=\left\{(x,v)\in{\cal H}:\delta(x,v)=\theta\right\}$.
It is clear that $\Gamma_0=\{(0,v):v\ge 0\}$, and moreover
\begin{equation}
\Gamma_{-\theta}=\{(x,v)\in{\cal H}:(-x,v)\in\Gamma_{\theta}\}.
\label{E:reflect}
\end{equation}

We will next study the structure of the level sets of $\delta$. It is not hard to see that the functions 
$$
\delta\mapsto\delta-\sin\delta\quad\mbox{and}\quad\delta\mapsto 2\sin\frac{\delta}{2}-\delta\cos\frac{\delta}{2}
$$ 
are positive and increasing on $(0,2\pi)$ (negative and increasing on $(-2\pi,0)$). It follows from (\ref{E:forma1}) 
that for $0<\theta< 2\pi$, the
level set $\Gamma_{\theta}$ is contained in the set ${(x,v)\in{\cal H}:x> 0}$. Moreover, it is not hard to see that the Heston manifold is covered by a disjoint family of the level sets $\Gamma_{\theta}$ with 
$-2\pi<\theta<2\pi$. Note that
for fixed $\theta$ with $0<\theta<2\pi$, the level curve $\Gamma_{\theta}$ intersects 
the half-line 
$\{(x,v)\in{\cal H}:x> 0,v=0\}$
at the point $(\psi(\theta),0)$ where the function $\psi$ is defined by
\begin{equation}
\psi(\theta)=\frac{\theta-\sin\theta}{1-\cos\theta}.
\label{E:psik}
\end{equation}
The previous statement follows from (\ref{E:forma1}). Therefore, for $0<\theta<2\pi$,
$$
\Gamma_{\theta}\subset\left\{(x,v)\in{\cal H}:x\ge\psi(\theta)\right\},
$$ 
and thus
\begin{equation}
x(1-\cos\theta)-(\theta-\sin\theta)\ge 0.
\label{E:there}
\end{equation}
\begin{lemma}\label{L:increas}
The function $\psi$ defined by (\ref{E:psik}) is strictly increasing on the interval $(0,2\pi)$.
\end{lemma}

\it Proof. \rm Differentiating the function $\psi$, we obtain
\begin{align*}
\psi^{\prime}(\theta)&=\frac{(1-\cos\theta)^2-(\theta-\sin\theta)\sin\theta}{(1-\cos\theta)^2} 
=\frac{2-2\cos\theta-\theta\sin\theta}{(1-\cos\theta)^2} \\
&=\frac{2\sin\frac{\theta}{2}\left(2\sin\frac{\theta}{2}-\theta\cos\frac{\theta}{2}\right)}
{(1-\cos\theta)^2}> 0.
\end{align*}

This completes the proof of Lemma \ref{L:increas}.

The function $\psi$ in (\ref{E:psik}) is strictly increasing on the interval $(0,2\pi)$. Moreover, $\psi$ maps  
$(0,2\pi)$ onto $(0,\infty)$. Therefore, the inverse function 
$\psi^{-1}$ exists on $(0,\infty)$, and the range of the function $\psi^{-1}$ is the interval $(0,2\pi)$.

We will next show that the level set $\Gamma_{\theta}$ with $0<\theta< 2\pi$ coincides with the graph of a certain function. To find such a characterization, we plug $\delta=\theta$ into the formula in (\ref{E:forma1}), and then solve the resulting quadratic equation for $\sqrt{v}$. This gives
\begin{equation}
\sqrt{v}=\frac{U(\theta,x)}{\theta-\sin\theta},
\label{E:clear}
\end{equation}
where
\begin{align}
&U(\theta,x)=\theta\cos\frac{\theta}{2}-2\sin\frac{\theta}{2}+\sqrt{N(\theta,x)}.
\label{E:solut}
\end{align}
The function $N$ in (\ref{E:solut}) is defined as follows:
\begin{align}
&N(\theta,x)=\left(\theta\cos\frac{\theta}{2}-2\sin\frac{\theta}{2}\right)^2 
+(\theta-\sin\theta)[x(1-\cos\theta)-(\theta-\sin\theta)].
\label{E:solots}
\end{align} 
Note that the positive sign is chosen in front of the square root in (\ref{E:solut}), because we assume that $\sqrt{v}\ge 0$. The second solution to the quadratic equation mentioned above is negative. Note also that since (\ref{E:there}) holds, the function under the square root sign in (\ref{E:solut})
is positive.
\begin{remark}\label{R:addi} \rm
For a fixed $\theta$ with $-2\pi<\theta< 0$, we choose the negative sign in front of the square root in (\ref{E:solut}). In this case, we have the following description of the level set $\Gamma_{\theta}$:
$$
\sqrt{v}=\frac{\widehat{U}(\theta,x)}{\theta-\sin\theta},
$$
where
$$
\widehat{U}(\theta,x)=\theta\cos\frac{\theta}{2}-2\sin\frac{\theta}{2}-\sqrt{N(\theta,x)},\quad x<\psi(\theta),
$$
and the function $N$ is defined by (\ref{E:solut}).
\end{remark}

It is easy to see that the following equalities hold:
\begin{align}
&N(\theta,x)=(\theta-\sin\theta)(1-\cos\theta)x+\left(\theta\cos\frac{\theta}{2}-2\sin\frac{\theta}{2}\right)^2
-(\theta-\sin\theta)^2 \nonumber \\
&=\sin^2\frac{\theta}{2}\left[2(\theta-\sin\theta)x+2(1-\cos\theta)-\theta^2\right].
\label{E:solot}
\end{align} 

The next lemma provides a description of the level sets with $0<\theta< 2\pi$.
\begin{lemma}\label{L:struc1}
For every $0<\theta< 2\pi$, 
\begin{equation}
\Gamma_{\theta}=\left\{(x,v)\in{\cal H}:
v=\frac{U(\theta,x)^2}{(\theta-\sin\theta)^2},\,\,\psi(\theta)\le x<\infty\right\}, 
\label{E:function1}
\end{equation}
where the function $U$ is given by (\ref{E:solut}). The function 
$x\mapsto v(\theta,x)$, $\psi(\theta)\le x<\infty$,
in (\ref{E:function1}) can be represented in the following form: 
\begin{equation}
v(\theta,x)=v_1(\theta,x)-v_2(\theta,x),
\label{E:repre}
\end{equation}
where 
\begin{equation}
v_1(\theta,x)=\frac{1-\cos\theta}{\theta-\sin\theta}x
+\frac{2\left(\theta\cos\frac{\theta}{2}-2\sin\frac{\theta}{2}\right)^2}{(\theta-\sin\theta)^2}
-1
\label{E:repr1}
\end{equation}
and
\begin{align}
&v_2(\theta,x)=\frac{2\sin\frac{\theta}{2}\left(2\sin\frac{\theta}{2}-\theta\cos\frac{\theta}{2}\right)}{(\theta-\sin\theta)^2}\sqrt{2(\theta-\sin\theta)x+2(1-\cos\theta)-\theta^2}.
\label{E:repr2}
\end{align}
The functions $v_1$ and $v_2$ in
(\ref{E:repr1}) and (\ref{E:repr2}) are nonnegative.
\end{lemma}
\begin{remark}\label{R:rr} \rm For a fixed $\theta$, the function $v_1$ in (\ref{E:repr1}) is an affine function,
while the function $v_2$ in (\ref{E:repr2}) is the square root of an affine function.
\end{remark}

\it Proof of Lemma \ref{L:struc1}. \rm It is not hard to see that (\ref{E:function1}) and (\ref{E:repre}) 
follow from (\ref{E:clear}), (\ref{E:solut}), and (\ref{E:solot}). The function $v_2$ is positive since
$\sin\frac{\theta}{2}> 0$ and $2\sin\frac{\theta}{2}-\theta\cos\frac{\theta}{2}> 0$ for $0<\theta< 2\pi$.
The positivity of the function $v_1$ can be established as follows. For $x>\psi(\theta)$, we have
$$
v_1(\theta,x)>\frac{2\left(\theta\cos\frac{\theta}{2}-2\sin\frac{\theta}{2}\right)^2}{(\theta-\sin\theta)^2}> 0.
$$

This completes the proof of Lemma \ref{L:struc1}.

Our next goal is to compute the derivatives of the function $v$ defined in (\ref{E:function1}).
Using (\ref{E:solut}), (\ref{E:solot}), and (\ref{E:function1}), we obtian
\begin{equation}
\frac{\partial v}{\partial x}=\frac{U(\theta,x)(1-\cos\theta)}{\sqrt{N(\theta,x)}(\theta-\sin\theta)},
\quad\psi(\theta)\le x<\infty,
\label{E:fff}
\end{equation}
In addition, using (\ref{E:solut}) in (\ref{E:fff}), we get
\begin{equation}
\frac{\partial v}{\partial x}=\frac{1-\cos\theta}{\theta-\sin\theta}\left\{1+
\frac{\theta\cos\frac{\theta}{2}-2\sin\frac{\theta}{2}}{\sqrt{N(\theta,x)}}\right\}
\label{E:rirr}
\end{equation}
for all $\psi(\theta)\le x<\infty$. Next, differentiating (\ref{E:rirr}), we obtain
\begin{equation}
\frac{\partial^2 v}{\partial x^2}=\frac{(2\sin\frac{\theta}{2}-\theta\cos\frac{\theta}{2})(1-\cos\theta)^2}{2N(\theta,x)^{\frac{3}{2}}}
\label{E:ror}
\end{equation}
for all $\psi(\theta)\le x<\infty$.
\begin{remark}\label{R:tan} \rm
Let $0<\theta< 2\pi$. Then the one-sided tangent line to the level curve $\Gamma_{\theta}$ at the point 
$(\psi(\theta),0)$ is horizontal. 
This follows from (\ref{E:fff}) and the equality $U(\theta,\psi(\theta))=0$.
\end{remark}

The next lemma describes the structural properties of the level sets $\Gamma_{\theta}$, $0<\theta< 2\pi$.
\begin{lemma}\label{L:struc2}
For every $0<\theta< 2\pi$, the function $x\mapsto v(\theta,x)$ defined in (\ref{E:function1})
is strictly increasing and convex on the interval $[\psi(\theta),\infty)$.
\end{lemma}
\begin{remark}\label{R:conve} \rm
For every $\theta$ with $-2\pi<\theta<2\pi$, the level set
$\Gamma_{\theta}$ is convex. Indeed, for $0<\theta< 2\pi$, the convexity of $\Gamma_{\theta}$ follows from 
Lemma \ref{L:struc2}. For $\theta=0$, we have $\Gamma_0=\{(0,v):v\ge 0\}$, while for $-2\pi<\theta< 0$, we can use
(\ref{E:reflect}).
\end{remark}

\it Proof of Lemma \ref{L:struc2}. \rm It is not hard to see, using (\ref{E:fff}) and (\ref{E:ror}),
that $\frac{\partial v}{\partial x}> 0$ and $\frac{\partial^2 v}{\partial x^2}> 0$ for all $\psi(\theta)\le x<\infty$.

This completes the proof of Lemma \ref{L:struc2}.

The next statement explains why the level sets of the function 
$\delta$ are important in the study of the distance to a line in the Heston manifold. 
\begin{theorem}\label{T:clear}
Let $-2\pi<\theta< 2\pi$, $\theta\neq 0$, and $(x,v)\in\Gamma_{\theta}$. 
Then
\begin{equation}
\frac{d_H((0,1),(x,v))^2}{2}=\Lambda(x,\theta),
\label{E:rec}
\end{equation}
where the function $\Lambda$ is defined by (\ref{E:recs}).
In addition, if $\theta=0$, then $x=0$, and
\begin{equation}
\frac{d_H((0,1),(0,v))^2}{2}=2(\sqrt{v}-1)^2.
\label{E:reec}
\end{equation}
\end{theorem}
\begin{remark}\label{R:important}
Theorem \ref{T:clear} will be used in the proof of Theorem \ref{T:kp}. We will only need to determine, which level sets $\Gamma_{\theta}$ intersect the given line, and then reduce the set of addmissible values of $\theta$ appropriately.
\end{remark}

\it Proof of Theorem \ref{T:clear}. \rm
If $0<\theta< 2\pi$, then formula (\ref{E:rec}) follows from (\ref{E:simple1}), (\ref{E:clear}), (\ref{E:solut}),
and (\ref{E:solot}). Next, suppose $-2\pi<\theta< 0$. Then formula (\ref{E:rec}) can be derived from the equality
\begin{equation}
d_H((0,1),(x,v))=d_H((0,1),(-x,v))
\label{E:sy}
\end{equation}
(use formula (\ref{E:synge}) in the proof). Finally, if $\theta=0$, then formula 
(\ref{E:reec}) can be obtained by passing to the limit as $\delta\rightarrow 0$ in formula (\ref{E:synge}).

The proof of Theorem \ref{T:clear} is thus completed.
\subsection{A new system of coordinates in the Heston manifold}\label{SS:news}
Using the parameter $\theta$, we can introduce a special system of coordinates
in the Heston manifold. 
\begin{definition}\label{D:newsy}
To any point $P\in{\cal H}$, we assign a label $(\theta,v)$ with $-2\pi<\theta< 2\pi$ and 
$v\ge 0$ as follows: The number $v$ is the second component 
of the point $P$ in the rectangular system of coordinates, while the number $\theta$ is the index of the unique level set $\Gamma_{\theta}$ such that $P\in\Gamma_{\theta}$. We will write $P=(\theta,v)$, and call the system of coordinates described above the $\delta$-system.
\end{definition}

It is clear that $P\leftrightarrow(\theta,v)$ is a one-to-one correspondence between ${\mathcal H}$ and $(-2\pi,2\pi)\times[0,\infty)$. It is also clear that if $P=(x,v)$
in the rectangular system of coordinates, then $P=(\theta,v)$ with $\theta=\delta((0,1),(x,v))$ in the $\delta$-system. One can compute the Jacobian determinant associated with the change of variables $(x,v)\mapsto(\theta,v)$, using the equation $f(v,\delta)=x$, where the function $f$ is defined by (\ref{E:forma1}). We leave these computations as an exercise for the interested reader. 

Formulas (\ref{E:synge}), (\ref{E:usef2}), and (\ref{E:simple1}) allow us to represent the Heston distance 
$d_H((0,1),P)$ in the $\delta$-system of coordinates. For instance, formula (\ref{E:usef2}) implies that
if $P=(\theta,v)$, then
\begin{equation}
d_H((0,1),P)=\frac{\theta}{\sin\frac{\theta}{2}}
\sqrt{(\sqrt{v}-1)^2+4\sqrt{v}\sin^2\frac{\theta}{4}}.
\label{E:useff}
\end{equation}

The next proposition states that the function $(\theta,v)\mapsto d_H((0,1),(\theta,v))$ is increasing componentwise 
on the set $[0,2\pi)\times[1,\infty)$. Note that a similar lemma does not hold for the Heston manifold equipped with the rectangular system of coordinates.
\begin{lemma}\label{L:incr}
Let $P_1=(\theta_1,v_1)$ and $P_2=(\theta_2,v_2)$ be points in the Heston manifold, and suppose $0\le\theta_1\le\theta_2< 2\pi$ and $1\le v_1\le v_2$. Then $d_H((0,1),P_1)\le d_H((0,1),P_2)$.
\end{lemma}

\it Proof. \rm Lemma \ref{L:incr} can be easily derived from formula (\ref{E:useff}). Note that the condition
$1\le v_1$ is important for the validity of Lemma \ref{L:incr}. Indeed, we can construct a counterexample in the case where $0\le v_1< 1$ as follows. Take $v_1=0$, $v_2=1$, and $\theta_1=\theta_2=\varepsilon$, where $0<\varepsilon<\pi$
and $2-2\cos\frac{\varepsilon}{2}< 1$. Then formula (\ref{E:synge}) implies that
$d_H((0,1),P_1)> d_H((0,1),P_2)$. 
\subsection{Distance to a level set of $\delta$}\label{SS:levelset}
Our main goal in the present section is to compute the following number:
$$
\widehat{D}_{\theta}=\min_{\left\{(x,v)\in\Gamma_{\theta}\right\}}\left\{d_H((0,1),(x,v))\right\},
$$
where $-2\pi<\theta<2\pi$. The number $D_{\theta}$ is the Heston distance from the point $(1,0)$ to the level 
set $\Gamma_{\theta}$ of the function $\delta$. Put 
$$
D_{\theta}=\min_{\left\{(x,v)\in\Gamma_{\theta}\right\}}\left\{\frac{d_H((0,1),(x,v))^2}{2}\right\}.
$$
It is clear that
\begin{equation}
\widehat{D}_{\theta}=\sqrt{2D_{\theta}}.
\label{E:ddd}
\end{equation}

We will use formulas (\ref{E:rec}) and (\ref{E:ddd}) to compute $\widehat{D}_{\theta}$. It follows from (\ref{E:rec}) that for $0<\theta< 2\pi$, we have
\begin{align}
&D_{\theta}=\min_{\left\{x:x\ge\psi(\theta)\right\}}\left\{\Lambda(x,\theta)\right\}, 
\label{E:minima1}
\end{align}
where the function $\psi$ and $\Lambda$ are defined by (\ref{E:psik}) and (\ref{E:recs}), respectively. To find the critical points corresponding to the minimization problem in (\ref{E:minima1}), we reduce the equation 
$\frac{\partial\Lambda}{\partial x}=0$ to the following:
$$
\frac{1-\cos\theta}{\sqrt{2x(\theta-\sin\theta)+2(1-\cos\theta)-\theta^2}}=1.
$$
Next, solving the previous equation, we see that the unique solution is given by
\begin{equation}
x_0(\theta)=\frac{\theta+\sin\theta}{2}.
\label{E:cri1}
\end{equation}

Our next goal is to explain when the critical point $x_0(\theta)$ given by (\ref{E:cri1}) belongs to the set $[\psi(\theta),\infty)$.
\begin{lemma}\label{L:critic}
Let $0<\theta< 2\pi$. Then the following statements hold:
\begin{enumerate}
\item If $\pi<\theta< 2\pi$, then $x_0(\theta)<\psi(\theta)$.
\item If $0<\theta<\pi$, then $x_0(\theta)> \psi(\theta)$.
\item If $\theta=\pi$, then $x_0(\theta)=\psi(\theta)$.
\end{enumerate}
\end{lemma}

\it Proof. \rm The validity of statement 3 in Lemma \ref{L:critic} is clear. We will next prove statement 1. If $\pi<\theta< 2\pi$,
then $\theta=\eta+\pi$, where $0<\eta<\pi$. It follows that
\begin{align*}
x_0(\theta)&=\frac{\theta+\sin\theta}{2}=\frac{\eta+\pi-\sin\eta}{2}<\frac{\eta+\pi+\sin\eta}{1+\cos\eta} \\
&=\frac{\theta-\sin\theta}{1-\cos\theta}=\psi(\theta).
\end{align*}
This establishes statement 1.

It remains to prove statement 2 in Lemma \ref{L:critic}. We will show that the following inequalities hold:
\begin{equation}
\psi(\theta)<\frac{\theta}{2}< x_0(\theta)\quad\mbox{for all}\quad 0<\theta<\pi.
\label{E:stat1}
\end{equation}
Indeed, the inequality $\frac{\theta}{2}< x_0(\theta)$ is straightforward. On the other hand, we have
$$
\psi(\theta)<\frac{\theta}{2}\Leftrightarrow\theta+\theta\cos\theta\le 2\sin\theta
\Leftrightarrow\theta\cos\frac{\theta}{2}< 2\sin\frac{\theta}{2}.
$$
The last inequality follows from the inequality 
$u<\tan u$, $0< u<\frac{\pi}{2}$.
This establishes (\ref{E:stat1}) and completes the proof of Lemma \ref{L:critic}.

The next proposition provides formulas for the distance to a level set of the function $\delta$ in the Heston manifold.
\begin{theorem}\label{T:ditol}
The following are true:
\begin{enumerate}
\item If $0\le|\theta|<\pi$, then 
\begin{equation}
\widehat{D}_{\theta}=|\theta|=d_H\left((0,1),\left(\frac{\theta+\sin\theta}{2},\cos^2\frac{\theta}{2}\right)\right).
\label{E:011}
\end{equation}
\item If $\pi\le|\theta|< 2\pi$, then
\begin{equation}
\widehat{D}_{\theta}=\theta\left(\sin\frac{\theta}{2}\right)^{-1}=d_H\left((0,1),\left(\frac{\theta-\sin\theta}{1-\cos\theta},0\right)\right).
\label{E:012}
\end{equation}
\end{enumerate}
\end{theorem}

\it Proof. \rm The equality $D_0=0$ is trivial. Moreover, it suffices to prove Theorem 
\ref{T:ditol} for $0<\theta<2\pi$, by the symmetry properties of the Heston distance. The following equalities can be obtained from (\ref{E:recs}) by direct computations:
\begin{equation}
\Lambda\left(x_0(\theta),\theta\right)=\Lambda\left(\frac{\theta+\sin\theta}{2},\theta\right)=\frac{\theta^2}{2},
\label{E:obs1}
\end{equation}
\begin{equation}
\Lambda\left(\psi(\theta),\theta\right)=\Lambda\left(\frac{\theta-\sin\theta}{1-\cos\theta},\theta\right)=\frac{\theta^2}{1-\cos\theta},
\label{E:obs2}
\end{equation}
and
\begin{equation}
\lim_{x\rightarrow\infty}\Lambda\left(x,\theta\right)=\infty.
\label{E:obs3}
\end{equation}
Note that the equality in (\ref{E:obs2}) can also be easily derived by plugging $x=\psi(\theta)$ and $v=0$ into (\ref{E:simple1}). In addition, (\ref{E:obs1}) and (\ref{E:obs2}) imply that
\begin{equation}
\Lambda\left(x_0(\theta),\theta\right)<\Lambda\left(\psi(\theta),\theta\right).
\label{E:obs4}
\end{equation}
Finally, it is not hard to see that the first equalities in (\ref{E:011}) and (\ref{E:012}) follow from (\ref{E:obs1}) - (\ref{E:obs4}), Lemma \ref{L:critic}, and (\ref{E:ddd}).

To prove the second equality in (\ref{E:011}), we have to compute the value of $v$ corresponding to $x=\frac{\theta+\sin\theta}{2}$ and $\theta$. Using (\ref{E:simple1}) and the first equality in (\ref{E:011}), 
we see that $v$ satisfies the following equation:
$$
\frac{\theta^2}{\theta-\sin\theta}\left(\frac{\theta+\sin\theta}{2}-2\sqrt{v}\sin\frac{\theta}{2}\right)
=\frac{\theta^2}{2}.
$$
Solving the previous equation, we obtain $v=\cos^2\frac{\theta}{2}$. This establishes the second equality
in (\ref{E:011}). The proof of the second equality in (\ref{E:012}) is straightforward.

This completes the proof of Theorem \ref{T:ditol}.

Theorem \ref{T:ditol} states that for $0\le\theta\le\pi$, the points where the distance from $(0,1)$ to 
$\Gamma_{\theta}$ is attained lie on the parametrized curve given by
\begin{equation}
\theta\mapsto P_{\theta}
=\left(\frac{\theta+\sin\theta}{2},\cos^2\frac{\theta}{2}\right),\quad 0\le\theta\le\pi.
\label{E:curve1}
\end{equation}
This curve is the graph of the following function:
\begin{equation}
v=\frac{1}{2}\left[1+\cos\left(x_0^{-1}(x)\right)\right],\quad 0\le x\le\frac{\pi}{2}.
\label{E:curve2}
\end{equation} 
It is clear that the curve connects the points $(0,1)$ and $(\frac{\pi}{2},0)$. 
\begin{lemma}\label{L:curve}
The function in (\ref{E:curve2}) is strictly decreasing and concave on the interval $[0,\frac{\pi}{2}]$.
\end{lemma}

\it Proof. \rm Since $x_0(\theta)=\frac{\theta+\sin\theta}{2}$, $\theta\in[0,\pi]$, we see that for all
$x\in[0,\frac{\pi}{2}]$,
$$
v^{\prime}(x)=-\frac{\sin(x_0^{-1}(x))}{1+\cos(x_0^{-1}(x))}\le 0
$$
and
\begin{align*}
v^{\prime\prime}(x)&=\frac{\sin(x_0^{-1}(x))
\cos(x_0^{-1}(x))^{\prime}-\sin(x_0^{-1}(x))^{\prime}(1+\cos(x_0^{-1}(x))}{(1+\cos(x_0^{-1}(x))^2} \\
&=-\frac{2}{(1+\cos(x_0^{-1}(x)))^2}\le 0.
\label{E:curve3}
\end{align*}

This completes the proof of Lemma \ref{L:curve}.
\begin{lemma}\label{L:curvet}
Every line $L_{\beta,\gamma}$ with $0\le\beta\le\frac{\pi}{2}$ and $\gamma\ge 0$ intersects the curve described
in (\ref{E:curve1}) at exactly one point $P_{\theta(\beta,\gamma)}$. Here the number $\theta(\beta,\gamma)$
is the unique solution to the equation $\zeta_{\gamma}(\theta)=\beta$, $0\le\theta\le\pi$, where
\begin{equation}
\zeta_{\gamma}(\theta)=\frac{\theta+\sin\theta-\gamma(1+\cos\theta)}{2}.
\label{E:solvab1}
\end{equation}
\end{lemma}

\it Proof. \rm The components $(x,v)$ of the intersection point satisfy $\beta+\gamma v=\frac{\theta+\sin\theta}
{2}$ and $v=\frac{1+\cos v}{2}$. Therefore, the intersecton point exists if and only if the equation in 
(\ref{E:solvab1}) is solvable. 
The function $\zeta_{\gamma}$ is strictly increasing on the interval $[0,\pi]$ 
and maps this interval onto the interval
$[-\gamma,\frac{\pi}{2}]$. Since $[0,\frac{\pi}{2}]\subset[-\gamma,\frac{\pi}{2}]$, Lemma \ref{L:curvet} holds.
\begin{remark}\label{R:2} \rm It is clear that $\theta(\beta,\gamma)=\zeta_{\gamma}^{-1}(\beta)$ and
$$
P_{\theta(\beta,\gamma)}=\left(\frac{\zeta_{\gamma}^{-1}(\beta)+\sin\zeta_{\gamma}^{-1}(\beta)}{2},\frac{1+\cos\zeta_{\gamma}^{-1}(\beta)}{2}\right).
$$
Note also that $\theta(0,\gamma)=\tau^{-1}(\gamma)$, where 
$$
\tau(\theta)=\frac{\theta+\sin\theta}{1+\cos\theta},\quad 0\le\theta<\pi.
$$
\end{remark}
\begin{remark}\label{R:cu} \rm For $\pi\le\theta< 2\pi$, the critical point is given by $P_{\theta}=(\psi(\theta),0)$. Therefore, 
it is natural to set $\theta(\beta,\gamma)=\psi^{-1}(\beta)$, for a line $L_{\beta,\gamma}$ with 
$\beta\ge\frac{\pi}{2}$ and $\gamma> 0$. 
\end{remark}
\subsection{Distance to a horizontal line in the Heston manifold}\label{SS:dhl}
The next assertion describes the behavior of the Heston distance function along a horizontal line.
\begin{lemma}\label{L:horiz}
Fix $v\ge 0$. Then the function $\rho(x)=d_H((0,1),(x,v))$ is strictly increasing on $[0,\infty)$.
\end{lemma}

\it Proof. \rm Using the level sets of $\delta$, we see that as the variable $x$ increases from $0$ to $\infty$ along the horizontal line, the corresponding function $\delta(x,v)$ increases from $0$ to $2\pi$.
We have
\begin{equation}
\left[\frac{\delta}{\sin\frac{\delta}{2}}\right]^{\prime}=\frac{\sin\frac{\delta}{2}-\frac{1}{2}\delta\cos\frac{\delta}
{2}}{\sin^2\frac{\delta}{2}}.
\label{E:hor}
\end{equation}
It is easy to see that the function in the numerator of the fraction on the right-hand side of (\ref{E:hor}) is positive. For $\pi<\delta< 2\pi$, this is clear, while for $0<\delta<\pi$, the previous statement follows from the
inequality $\tan u> u$, $0< u<\frac{\pi}{2}$.
Therefore, the function 
$$
\delta\mapsto\frac{\delta^2}{2}\left(\sin\frac{\delta}{2}\right)^{-1}
$$
increases on the interval $(0,2\pi)$. In addition, the function 
$\delta\mapsto-\cos\frac{\delta}{2}$
increases on $(0,2\pi)$. Now Lemma \ref{L:horiz} follows from formula (\ref{E:synge}).

Let us fix $\tau\ge 0$, and denote by $\rho_{\tau}$ the horizontal line defined by $\rho_{\tau}=
\{(x,v)\in{\cal H}:v=\tau\}$. The next assertion provides a formula for the distance from a point $(0,1)$ 
to the line $\rho_{\tau}$. Note that $\rho_{\tau}$ is a level curve with respect to the second component in the $\delta$-system of coordinates.
\begin{theorem}\label{T:conclude}
For all $\tau\ge 0$, $d_H((0,1),\rho_{\tau})=2|\sqrt{\tau}-1|$.
\end{theorem}

Theorem \ref{T:conclude} follows from Lemma \ref{L:horiz} and formula
(\ref{E:reec}).
\subsection{Minimization problems and the proof of Theorem \ref{T:kp}}\label{SS:dttl}
In the present subsection, we show that the number $D_{\beta,0}$, corresponding to a vertical line in the
Heston manifold, is the solution to a minimization problem for a certain function of one variable.

We will next characterize the limiting behavior of the Heston distance on the vertical line 
$L_{\beta,0}$.
\begin{lemma}\label{L:lim}
Let $\beta\in\mathbb{R}$. Then the following equality holds
\begin{equation}
\lim_{v\rightarrow\infty}\frac{d_H((0,1),(\beta,v))}{\sqrt{v}}=2.
\label{E:suffiks}
\end{equation}
Therefore, for every $\beta\in\mathbb{R}$,
$$
\lim_{v\rightarrow\infty}\frac{d_H((0,1),(\beta,v))}{d_H((0,1),(0,v))}=1.
$$
\end{lemma}

\it Proof. \rm Using the symmetry properties of the Heston distance, we see that it suffices to prove 
(\ref{E:suffiks}) for $\beta\ge 0$. For $\beta=0$, the equality in (\ref{E:suffiks}) can be easily derived from (\ref{E:reec}).
Next, let $\beta> 0$. It is not hard to see, using the level sets of $\delta$, that if $x=\beta$ and $v\rightarrow\infty$, then $\delta\rightarrow 0$. Now (\ref{E:suffiks}) follows from (\ref{E:usef2}).

This completes the proof of Lemma \ref{L:lim}.

It is obvious that $D_{0,0}=0$. With no loss of generality, we can restrict ourselves to the case where $\beta> 0$, since $D_{-\beta,-\gamma}=D_{\beta,\gamma}$. The previous equality follows from (\ref{E:sy}).

The next assetrion reduces the problem of computing the number $D_{\beta,0}$ to a minimization problem for a function of one variable.
\begin{theorem}\label{T:min}
Suppose $\beta> 0$. Then the following equality holds:
\begin{align}
&D_{\beta,0}=\inf_{\theta\in(0,\psi^{-1}(\beta)]}\left\{\Lambda(\beta,\theta)\right\},
\label{E:record}
\end{align}
where the function $\Lambda$ is defined by (\ref{E:recs}).
\end{theorem}

\it Proof. \rm The line $L_{\beta,0}$ intersects the level set $\Gamma_{\theta}$ 
with $0<\theta< 2\pi$ no more than once. Set
$S_{\beta}=\left\{\theta\in(0,2\pi):L_{\beta,0}\cap\Gamma_{\theta}\neq\emptyset\right\}$.
It is not hard to see that 
$S_{\beta}=\left\{\theta:0<x_{\theta}\le\beta\right\}=(0,\psi^{-1}(\beta)]$.
Now formula (\ref{E:record}) follows from (\ref{E:rec}).

This completes the proof of Theorem \ref{T:min}.

The level curves $\Gamma_{\theta}$ with $\theta=\pi$ and $\theta=-\pi$ play an especially important
role in the $\delta$-analysis of the Heston geometry. Note that we have $\psi(\pi)=\frac{\pi}{2}$ 
and $\psi(-\pi)=-\frac{\pi}{2}$. For the fixed initial point $(0,1)$, we call the subset of the Heston manifold ${\cal H}$, bounded by the level set $\Gamma_{-\pi}$, the segment $[-\frac{\pi}{2},
\frac{\pi}{2}]$, and the level set $\Gamma_{\pi}$, the set of $\delta$-close points with respect 
to the point $(0,1)$, while the complement of this set in ${\cal H}$ is called the set of 
$\delta$-far points
with respect to $(0,1)$ (see \cite{GL}). In terms of the paramter $\delta$, the close-point regime is characterized by $\delta\in[-\pi,\pi]$, while the far point regime is described by 
$\delta\in(-2\pi,-\pi)\cup(\pi,2\pi)$.

Suppose $0<\beta\le\frac{\pi}{2}$. Then the whole line $L_{\beta,0}$ is contained in the set of $\delta$-close points with respect to the point $(0,1)$. However, if $\frac{\pi}{2}<\beta<\infty$, then 
the initial segment of the line $L_{\beta,0}$ (the set of the points on $L_{\beta,0}$ for which $\pi<\delta\le\psi^{-1}(\beta))$) 
is contained in the set of $\delta$-far points, while the rest of the line $L_{\beta,0}$ (the set of the points on $L_{\beta,0}$ for which $0<\delta
\le\pi$) is contained in the set of $\delta$-close points with respect to the point $(0,1)$. We will
next show that the initial segment can be disregarded in formula (\ref{E:record}). This simplifies computations of the number $D_{\beta,0}$.
\begin{corollary}\label{C:simplifies}
Suppose $\beta\ge\frac{\pi}{2}$. Then the following equality holds:
\begin{align}
&D_{\beta,0}=\inf_{\theta\in(0,\pi]}\left\{\Lambda(\beta,\theta)\right\}.
\label{E:alcases}
\end{align}
\end{corollary}

\it Proof. \rm The proof is based on formula (\ref{E:simple1}). Set
$\xi(\delta)=\frac{\delta^2}{\delta-\sin\delta}$, and suppose $0<\delta< 2\pi$. Then we have
\begin{align*}
\xi^{\prime}(\delta)&=\frac{\delta}{(\delta-\sin\delta)^2}(\delta-2\sin\delta+\delta\cos\delta)
=\frac{2\delta\cos\frac{\delta}{2}}{(\delta-\sin\delta)^2}\left(\delta\cos\frac{\delta}{2}
-2\sin\frac{\delta}{2}\right).
\end{align*}
Now, it is easy to see that $\xi^{\prime}(\delta)< 0$ if $0<\delta<\pi$, $\xi^{\prime}(\delta)> 0$ if $\pi<\delta
<2\pi$, and $\xi^{\prime}(\pi)=0$. Therefore, the function $\xi$ is strictly increasing on the interval $[\pi,2\pi)$ 
and strictly decreasing on the interval $(0,\pi]$.

Now imagine that we travel along the line $L_{\beta,0}$ from the point $(\beta,0)$ 
(we have $\delta=\psi^{-1}(\beta)$ for this point) to the point on the border between the set of $\delta$-close points and the set of $\delta$-far points ($\delta=\pi$ there). Then $\delta$ decreases, and taking into account the previous reasoning, we see that the function $\xi$ also decreases. Moreover, the function
$(\delta,v)\mapsto\sqrt{v}\sin\frac{\delta}{2}$ increases. Overall, the expression on the right-hand side of formula (\ref{E:simple1}) decreases during such a trip. It follows that the initial segment of the line
$L_{\beta,0}$, where $\psi^{-1}(\beta)\le\delta<\pi$, does not provide any input into formula (\ref{E:rec}).

This completes the proof of Corollary \ref{C:simplifies}.
\begin{lemma}\label{L:1}
The minimization problems in (\ref{E:record}) and (\ref{E:alcases}) admit a unique solution.
\end{lemma}

\it Proof. \rm
Using the definition in (\ref{E:distance}), we see that for $0<\beta<\frac{\pi}{2}$,
\begin{equation}
D_{\beta,0}=\inf_{\{v:v\ge 0\}}\left\{\frac{d_H((0,1),(\beta,v))^2}{2}\right\}.
\label{E:stance1}
\end{equation}
Moreover, Corollary \ref{C:simplifies} implies that for $\frac{\pi}{2}\le\beta<\infty$,
\begin{equation}
D_{\beta,0}=\inf_{\{v:v\ge\tilde{v}(\beta)\}}\left\{\frac{d_H((0,1),(\beta,v))^2}{2}\right\},
\label{E:stance2}
\end{equation}
where 
\begin{equation}
\tilde{v}(\beta)=v(\beta,\pi)=\frac{8+2\pi\beta-\pi^2-4\sqrt{4+2\pi\beta-\pi^2}}{\pi^2}.
\label{E:stance3}
\end{equation}
The second equality in (\ref{E:stance3}) follows from (\ref{E:forma1}). Note that the minimization problems in 
(\ref{E:stance1}) and (\ref{E:stance2}) correspond to the points on the line $L_{\beta,0}$ in the close-point regime
with respect to the point $(0,1)$.

Our next goal is to show that the minimization problems in 
(\ref{E:stance1}) and (\ref{E:stance2}) have a unique solution. It is not hard to check that the infimum in (\ref{E:stance1}) and (\ref{E:stance2}) can not be at infinity. We will next formulate several results from \cite{GL},
which will be needed in the proof. 
The following equality is valid:
\begin{equation}
d_H((0,1),(x,y^2))=d_G((0,1),(x,y)),\quad x\in\mathbb{R},\quad y> 0,
\label{E:mp}
\end{equation}
where $d_G$ is the Carnot-Carath\'{e}odory distance in the Grushin model associated with the Heston model (see
formula (30) in \cite{GL}). The function 
$$
\widetilde{\Lambda}(x,y)=\frac{d^2_G((0,1),(x,y))}{2}
$$
is strictly convex on the set $\widetilde{S}$ of all the points in the Grushin upper half-plane, which are $\delta$-close points with respect to $(0,1)$. This follows from the fact that the function $\widetilde{\Lambda}$ coincides with 
the Legendre-Fenchel transform of the limiting cumulant generating function for the Grushin model (see Theorem 23
in \cite{GL}). It follows from (\ref{E:mp}) that the problem of finding the infimum of the function 
$y\mapsto\frac{d_G^2((0,1),(\beta,y))}{2}$ on the intersection of the line $L_{\beta,0}$ with the set $\widetilde{S}$
is a convex minimization problem, and therefore it has a unique minimum point $y^{*}(\beta)\ge 0$. Now, (\ref{E:mp}) implies that $y^{\star}(\beta)$ is the unique point minimizing the function $y\mapsto\frac{d_H^2((0,1),(x,y^2))}{2}$  on the set $L_{\beta,0}\cap\widetilde{S}$. The transformation $(x,y)\mapsto(x,v)$, where $v=y^2$, is a one-to-one mapping of the set $\widetilde{S}$ onto the set of all the points in the Heston upper half-plane, which are $\delta$-close points with respect to $(0,1)$. Therefore, the point $v^{*}(\beta)=y^{\star}(\beta)^2$ is the unique point minimizing the function $y\mapsto\frac{d_H^2((0,1),(x,v))}{2}$  on the set $L_{\beta,0}\cap S$. 

The proof of Lemma \ref{L:1} is thus completed.

For $\frac{\pi}{2}<\beta<\infty$, Corollary \ref{C:simplifies} allows us to reduce the domain $(0,\psi^{-1}(\beta)]$ in the minimization problem $\inf_{\theta\in(0,\psi^{-1}(\beta)]}\left\{\Lambda(\beta,\theta)\right\}$ to a smaller domain $(0,\pi]$. We will next make a similar reduction in the case where $0<\beta<\frac{\pi}{2}$. Moreover, we will replace the domains in both minimization problems described above by smaller closed intervals.

The following notation will be used in the next corollary. Recall that we denoted by 
$x_0$ the function defined by 
$x_0(\theta)=\frac{\theta+\sin\theta}{2}$
(see (\ref{E:cri1})). The function $x_0$ is strictly increasing on the interval $0\le\theta\le\pi$ and maps this interval onto the interval $[0,\frac{\pi}{2}]$. Therefore, the inverse function $x_0^{-1}$ is defined on the interval $[0,\frac{\pi}{2}]$. Set
\begin{equation}
\tau(\beta)=\left(\frac{x_0^{-1}(\beta)}{2}+1\right)^2,\quad 0<\beta<\frac{\pi}{2}.
\label{E:wh}
\end{equation}

Throughout the rest of the paper, we will denote by $\delta(x,v)$ the number $\delta((0,1),(x,v))$.
\begin{corollary}\label{C:reduction1}
If $0<\beta<\frac{\pi}{2}$, then
\begin{equation}
D_{\beta,0}=\min_{\{\theta:\delta(\beta,\tau(\beta))\le\theta\le x_0^{-1}(\beta)\}}
\left\{\Lambda(\beta,\theta)\right\}.
\label{E:wh1}
\end{equation}
\end{corollary}

\it Proof. \rm Let $0<\beta<\frac{\pi}{2}$, and put $\theta_0=x_0^{-1}(\beta)$. Then $\beta=\frac{\theta_0
+\sin\theta_0}{2}$, and moreover the minimum point 
$$
p=\left(\frac{\theta_0+\sin\theta_0}{2},\cos^2\frac{\theta_0}{2}\right)
$$
in (\ref{E:011}) lies on the line $L_{\beta,0}$. It is not hard to see that the level set $\Gamma_{x_0}$ separates the 
point $(0,1)$ from the vertical segment 
$$
S=\left\{(x,y):x=\beta,0\le y\le\cos^2\frac{\theta_0}{2}\right\}.
$$
It follows from the definition of the Riemannian distance and (\ref{E:011}) that
$$
d_H((0,1);S)\ge\widehat{D}_{\theta_0}=d_H((0,1),p),
$$ 
and since the point $p$ belongs to the segment $S$, we have
$d_H((0,1);S)=d_H((0,1),p)$. This explains why the upper bound $\theta\le x_0^{-1}(\beta)$ can be used in formula
(\ref{E:wh1}).

We will next explain the appearance of the lower bound 
$\delta(\beta,\tau(\beta))\le\theta$
in formulas (\ref{E:wh1}) and (\ref{E:wh2}). Let us consider the horizontal line 
$\rho_{\tau(\beta)}=\{(x,v)\in{\cal H}:v=\tau(\beta)\}$,
where 
$\tau(\beta)$ is given by (\ref{E:wh}). This line separates the point $(0,1)$ from the set
$\widetilde{S}=\{(x,y):x=\beta,\tau(\beta)\le y<\infty\}$.
Therefore,  
$d_H((0,1);S)\ge d_H\left((0,1);\rho_{\tau(\beta)}\right)$.
Moreover, Theorem \ref{T:conclude}, (\ref{E:011}), and (\ref{E:wh}) imply that for $0<\beta<\frac{\pi}{2}$, 
$$
d_H\left((0,1);\rho_{\tau(\beta)}\right)=x_0^{-1}(\beta)=\theta_0=d_H((0,1),p).
$$
It follows that $d_H((0,1);S)\ge d_H((0,1),p)$.
Since $p\in L_{\beta,0}$, the numbers $\theta$ in (\ref{E:forma1}), corresponding to the points in the set $S$, that is, the numbers $\theta$ such that
$0<\theta<\delta(\beta,\tau(\beta))$,  can be disregarded in the minimization problem 
$$
\inf_{\{\theta:0<\theta\le x_0^{-1}(\beta)\}}\left\{\Lambda(\beta,\theta)\right\}.
$$ 
This establishes formula (\ref{E:wh1}) and completes the proof of Corollary \ref{C:reduction1}.

For $\frac{\pi}{2}\le\beta<\infty$, we consider the point $q\in L_{\beta,0}$ such that
$x=\beta$ and $\theta=\pi$ instead of the point $p$ used in the case where $0<\beta<\frac{\pi}{2}$. Note that for the point $q$ we have $v=\tilde{v}(\beta)$, where $\tilde{v}$ is given by (\ref{E:stance3}). Let us set
\begin{equation}
z(\beta)=\sqrt{2\pi\beta+8-4\sqrt{2\pi\beta+4-\pi^2}}.
\label{E:zb}
\end{equation}
Then, it is not hard to see, using 
formula (\ref{E:stance3}), that
$d_H((0,1),q)=z(\beta)$. 
We will also need the following function:
$$
\hat{\tau}(\beta)=\left(\frac{z(\beta)}{2}+1\right)^2.
$$
\begin{corollary}\label{C:reduction2}
If $\frac{\pi}{2}\le\beta<\infty$, then
\begin{equation}
D_{\beta,0}=\min_{\{\theta:\delta(\beta,\hat{\tau}(\beta))\le\theta\le\pi\}}
\left\{\Lambda(\beta,\theta)\right\}.
\label{E:wh2}
\end{equation}
\end{corollary}

The proof of Corollary \ref{C:reduction2} is similar to that of the corresponding part of Corollary \ref{C:reduction1},
and we leave it as an exercise for the interested reader.

Our final goal in the present section is to pass to larger intervals in the minimization problems in 
(\ref{E:wh1}) and (\ref{E:wh2}) in order to make the expressions for the end points of the minimization intervals simpler. This 
may be useful in numerical computations. 
\begin{corollary}\label{C:finnal}
If $0<\beta<\frac{\pi}{2}$, then
\begin{equation}
D_{\beta,0}=\min_{\{\theta:\delta\left(\beta,\left(\beta+1\right)^2\right)\le\theta\le 2\beta\}}
\left\{\Lambda(\beta,\theta)\right\},
\label{E:who1}
\end{equation}
while if $\frac{\pi}{2}\le\beta<\infty$, then
\begin{equation}
D_{\beta,0}=\min_{\{\theta:\delta(\beta,5\beta)\le\theta\le\pi\}}
\left\{\Lambda(\beta,\theta)\right\}
\label{E:who2}
\end{equation}
\end{corollary}

\it Proof. \rm Our first goal is to find simple estimates from above for the functions $\tau$ and $\hat{\tau}$.
It follows from (\ref{E:stat1}) that for $0<\beta<\frac{\pi}{2}$, 
\begin{equation}
x_0^{-1}(\beta)< 2\beta<\psi^{-1}(\beta),
\label{E:xx}
\end{equation} 
and hence
\begin{equation}
\tau(\beta)\le\left(\beta+1\right)^2.
\label{E:wbu2}
\end{equation}
Moreover, for $\frac{\pi}{2}\le\beta<\infty$, (\ref{E:zb}) implies that
$z(\beta)\le\sqrt{2\pi\beta}$. Hence,
\begin{equation}
\hat{\tau}(\beta)\le\left(\frac{\sqrt{2\pi\beta}}{2}+1\right)^2
\le\frac{\pi\beta}{2}+\sqrt{2\pi}\beta+\frac{2}{\pi}\beta\le 5\beta.
\label{E:wbu3}
\end{equation}

Analyzing the proofs of Corollaries \ref{C:reduction1} and \ref{C:reduction2}, we see that any numbers greater than  $\tau(\beta)$ and $\hat{\tau}(\beta)$ can be used in formulas (\ref{E:wh1}) and (\ref{E:wh2}). Moreover, we can take any number between $x_0^{-1}(\beta)$ and $\psi^{-1}(\beta)$ in formula (\ref{E:wh1}).
Finally, taking into account (\ref{E:xx}), (\ref{E:wbu2}), (\ref{E:wbu3}), and the previous observation, we establish 
Corollary \ref{C:finnal}.

\it Proof of Theorem \ref{T:kp}. \rm We will derive Theorem \ref{T:kp} from Corollary \ref{C:finnal}, using the estimates in Lemma \ref{L:major}. However, in order the proof given below to be correct, we need to show that 
for $0<\beta<\frac{\pi}{2}$, $[\frac{1}{11}\beta,2\beta]\subset(0,\psi^{-1}(\beta)]$ (see (\ref{E:record})), and for $\frac{\pi}{2}\le\beta<\infty$, $[\frac{1}{7},\pi]\subset(0,\pi]$ (see (\ref{E:alcases})). The latter inclusion is trivial, while the former one follows from the second inequality in (\ref{E:xx}).

For $0<\beta<\frac{\pi}{2}$, we have $0<\beta\le\frac{\pi}{2}\le\frac{\pi^2}{12}(v+\sqrt{v}+1)$ for any $v\ge 0$.
Therefore, we can apply the first estimate in (\ref{E:kabel}), which gives
\begin{align*}
&\delta\left(\beta,(\beta+1)^2\right)\ge\frac{12\beta}{\pi^2\left((\beta+1)^2+\beta+1+1\right)}
\ge\frac{12\beta}{10(\beta+2)^2}
\ge\frac{6\beta}{5\left(\frac{\pi}{2}+2\right)^2}\ge\frac{1}{11}\beta.
\end{align*}
Now, (\ref{E:whos1}) follows from (\ref{E:who1}).

Next, let $\frac{\pi}{2}\le\beta<\infty$. In this case, we need to estimate the quantity $\delta(\beta,5\beta)$ 
from below. We have $0<\beta\le\frac{\pi^3}{12}(5\beta+\sqrt{5\beta}+1)$. Applying the first estimate in
(\ref{E:kabel}), we obtain
\begin{align*}
\delta(\beta,5\beta)\ge\frac{12\beta}{\pi^2(5\beta+\sqrt{5\beta}+1)}\ge\frac{12\beta}
{10\left(5+\sqrt{5}+\frac{2}{\pi}\right)\beta}\ge\frac{1}{7}.
\end{align*}
It follows from the previous estimate and (\ref{E:who2}) that (\ref{E:whos2}) holds.

This completes the proof of Theorem \ref{T:kp}.
\section{Distance to a slanted line in the Heston manifold}\label{S:skew}
In this section, we turn our attention to the problem of computing the distance to the line $L_{\beta,\gamma}$ with $\gamma\neq 0$. It will be assumed below that $\beta\ge 0$, since the case where $\beta< 0$ can be dealt with, using the results obtained for $\beta> 0$ and the symmetry properties of the Heston distance. We will first suppose that
$\beta\ge 0$ and $\gamma> 0$. The case where $\beta> 0$ and $\gamma< 0$ will be considered later.

Let $\beta\ge 0$ and $\gamma> 0$. Then, a level set $\Gamma_{\theta}$ intersects the line $L_{\beta,\gamma}$ only if $0\le\theta<2\pi$. There is a crucial difference between the cases of vertical lines and slanted ones. For a vertical line, any level set $\Gamma_{\theta}$ intersects the line no more than once. On the other hand, a slanted line may have two intersection points with a level set $\Gamma_{\theta}$. In the latter case, there are two values of the parameter $v$ for which $(\beta+\gamma v, v)\in\Gamma_{\theta}$. 

Plugging $x=\beta+\gamma v$ into (\ref{E:clear}) and (\ref{E:solut}), we obtain the following equation with respect to $v$:
\begin{equation}
\sqrt{v}=A(\theta)+\sqrt{\gamma B(\theta)v+A(\theta)^2+\beta B(\theta)-1},
\label{E:gen1}
\end{equation}
where
\begin{equation}
A(\theta)=\frac{\theta\cos\frac{\theta}{2}-2\sin\frac{\theta}{2}}{\theta-\sin\theta}
\label{E:gen2}
\end{equation}
and
\begin{equation}
B(\theta)=\frac{1-\cos\theta}{\theta-\sin\theta}.
\label{E:gen3}
\end{equation}
In (\ref{E:gen1}), we should also assume that 
\begin{equation}
\beta+\gamma v\ge\psi(\theta),
\label{E:gg}
\end{equation}
since $x\ge\psi(\theta)$ in the definition of the level sets. It is easy to see that the inequality in 
(\ref{E:gg}) implies that the expression under the square root sign in (\ref{E:gen1}) is nonnegative.

It is clear that the equation in (\ref{E:gen1}) is equivalent to the following quadratic equation with respect to 
$\sqrt{v}$:
\begin{equation}
\left(1-\gamma B(\theta)\right)v-2A(\theta)\sqrt{v}+1-\beta B(\theta)=0.
\label{E:neo1}
\end{equation}
Equation (\ref{E:neo1}) allows us to represent $\sqrt{v}$ and $v$ as functions of $\theta$. It is easy to see, by solving equation (\ref{E:neo1}) for $\sqrt{v}$, that the following statement holds.
\begin{lemma}\label{L:vsqv}
If $1-\gamma B(\theta)\neq 0$, then the solutions to (\ref{E:neo1}) are as follows:
$\sqrt{v}=S^{+}_{\beta,\gamma}(\theta)$ and 
$\sqrt{v}=S^{-}_{\beta,\gamma}(\theta)$, where
\begin{equation}
S^{+}_{\beta,\gamma}(\theta)=\frac{A(\theta)+\sqrt{A(\theta)^2-(1-\gamma B(\theta))
(1-\beta B(\theta))}}{1-\gamma B(\theta)}
\label{E:S1}
\end{equation}
and
\begin{equation}
S^{-}_{\beta,\gamma}(\theta)=\frac{A(\theta)-\sqrt{A(\theta)^2-(1-\gamma B(\theta))
(1-\beta B(\theta))}}{1-\gamma B(\theta)}.
\label{E:S2}
\end{equation}
On the other hand,
if $1-\gamma B(\theta)=0$, then there is only one solution given by $\sqrt{v}=S_{\beta}(\theta)$ with
\begin{equation}
S_{\beta}(\theta)=\frac{1-\beta B(\theta)}{2A(\theta)}.
\label{E:S}
\end{equation}
\end{lemma}
\begin{remark}\label{R:syn} \rm
It follows from (\ref{E:S1}), (\ref{E:S2}), and (\ref{E:S}) that 
\begin{align}
S^{+}_{\beta,\gamma}(\theta)^2=
&\frac{1}{(1-\gamma B(\theta))^2}[2A(\theta)^2-(1-\gamma B(\theta))(1-\beta B(\theta)) \nonumber \\
&\quad+2A(\theta)
\sqrt{A(\theta)^2-(1-\gamma B(\theta))(1-\beta B(\theta))}],
\label{E:S12}
\end{align}
\begin{align}
S^{-}_{\beta,\gamma}(\theta)^2=
&\frac{1}{(1-\gamma B(\theta))^2}[2A(\theta)^2-(1-\gamma B(\theta))(1-\beta B(\theta)) \nonumber \\
&\quad-2A(\theta)
\sqrt{A(\theta)^2-(1-\gamma B(\theta))(1-\beta B(\theta))}].
\label{E:S22}
\end{align}
and
$$
S_{\beta}(\theta)^2=\frac{(1-\beta B(\theta))^2}{4A(\theta)^2}.
$$
\end{remark}

In our analysis of the Heston distance to a vertical line $L_{\beta,0}$, we used the function $\theta\mapsto\Lambda(\beta,\theta)$ (see Theorem \ref{T:clear}). Since for a slanted line there may be two choices for $\sqrt{v}$, two functions will be used in the minimization problems. Put
\begin{equation}
\Lambda^{+}_{\beta,\gamma}(\theta)=\frac{\theta^2}{1-\cos\theta}\left(S^{+}_{\beta,\gamma}(\theta)^2+1
-2S^{+}_{\beta,\gamma}(\theta)\cos\frac{\theta}{2}\right)
\label{E:anno1}
\end{equation}
and
\begin{equation}
\Lambda^{-}_{\beta,\gamma}(\theta)=\frac{\theta^2}{1-\cos\theta}\left(S^{-}_{\beta,\gamma}(\theta)^2+1
-2S^{-}_{\beta,\gamma}(\theta)\cos\frac{\theta}{2}\right),
\label{E:anno2}
\end{equation}
where the functions $S^{+}_{\beta,\gamma}$ and $S^{-}_{\beta,\gamma}$ are defined by
(\ref{E:S1}) and (\ref{E:S2}), respectively.  The definitions of the functions in (\ref{E:anno1}) and (\ref{E:anno2}), 
are based on formula (\ref{E:synge}).
\begin{lemma}\label{L:mono}
1.\,\,The function $B$ defined by (\ref{E:gen3}) is positive and strictly decreasing on the interval $(0,2\pi)$. 

2.\,\,The function $-A$, where $A$ is given by (\ref{E:gen2}), is positive and strictly increasing on the interval $(0,2\pi)$.
\end{lemma}

\it Proof. \rm The positivity statements in parts 1 and 2 are easy to establish. Since 
$B(\theta)=\frac{1}{\psi(\theta)}$, and Lemma \ref{L:increas} holds, the function $B$ is 
strictly decreasing.
 
We will next prove the monotonicity statement in part 2 of Lemma \ref{L:mono}. We have
$$
-A^{\prime}(\theta)=\frac{\frac{1}{2}\theta^2+\frac{1}{2}\theta\sin\theta-4\sin^2\frac{\theta}{2}}
{(\theta-\sin\theta)^2}.
$$
In order to prove that $-A^{\prime}(\theta)> 0$, it suffices to show that
\begin{equation}
\theta^2+\theta\sin\theta> 8\sin^2\frac{\theta}{2}.
\label{E:compare}
\end{equation}
Denote by $\lambda_1$ and $\lambda_2$ the functions on the left-hand side and the right-hand side
of (\ref{E:compare}), respectively. We have $\lambda_1(0)=\lambda_2(0)=0$, 
$\lambda_1^{\prime}(\theta)=2\theta+\sin\theta+\theta\cos\theta$, and
$\lambda_2^{\prime}(\theta)=4\sin\theta$.
Moreover, $\lambda_1^{\prime}(0)=\lambda_2^{\prime}(0)=0$,
$\lambda_1^{\prime\prime}(\theta)=2+2\cos\theta-\theta\sin\theta$, and
$\lambda_2^{\prime\prime}(\theta)=4\cos\theta$.
It follows that
\begin{align*}
&\lambda_1^{\prime\prime}(\theta)-\lambda_2^{\prime\prime}(\theta)=
4\sin^2\frac{\theta}{2}-2\theta\sin\frac{\theta}{2}\cos\frac{\theta}{2}
=2\sin\frac{\theta}{2}\left(2\sin\frac{\theta}{2}-\theta\cos\frac{\theta}{2}\right)> 0.
\end{align*}
Now we see that 
$\lambda_1^{\prime\prime}(\theta)-\lambda_2^{\prime\prime}(\theta)> 0\Rightarrow
\lambda_1^{\prime}(\theta)-\lambda_2^{\prime}(\theta)> 0\Rightarrow 
\lambda_1(\theta)-\lambda_2(\theta)> 0$.
This establishes inequality (\ref{E:compare}).

The proof of Lemma \ref{L:mono} is thus completed.
\begin{remark}\label{R:drugoj}
We define the number $S^{+}_{\beta,\gamma}(\psi^{-1}(\gamma))$ as follows:  
$$
S^{+}_{\beta,\gamma}(\psi^{-1}(\gamma))=\lim_{\theta\uparrow\psi^{-1}(\gamma)}S^{+}_{\beta,\gamma}(\theta)
$$
(the number 
$\Lambda^{+}_{\beta,\gamma}(\psi^{-1}(\gamma))$ is defined similarly). It is easy to see that $S^{+}_{\beta,\gamma}(\psi^{-1}(\gamma))=S_{\beta}(\psi^{-1}(\gamma))$, where the function
$S_{\beta}$ is given by (\ref{E:S}). On the other hand, the limit $\displaystyle{\lim_{\theta\uparrow\psi^{-1}(\gamma)}S^{+}_{\beta,\gamma}(\theta)}$ is not finite, and $\Lambda^{-}_{\beta,\gamma}(\psi^{-1}(\gamma))$ does not exist.
\end{remark}

In the sequel, we will use only those solutions to the equation in (\ref{E:gen1}), which are real and nonnegative. 
Let us assume that $0<\theta< 2\pi$ and $\psi(\theta)\neq\gamma$. Then the solutions are real if and only if 
\begin{equation}
A(\theta)^2-(1-\gamma B(\theta))(1-\beta B(\theta))\ge 0.
\label{E:za}
\end{equation}
Moreover, we can restrict ourselves to the case where $\psi(\theta)\le\max(\beta,\gamma)$, since otherwise we have
$S^{+}_{\beta,\gamma}(\theta)< 0$ and $S^{-}_{\beta,\gamma}(\theta)< 0$. It is easy to see that the inequality
in (\ref{E:za}) holds if and only if
\begin{equation}
\psi(\theta)^2A(\theta)^2-(\psi(\theta)-\gamma)(\psi(\theta)-\beta)\ge 0.
\label{E:eqqo}
\end{equation}

Suppose $\beta=\gamma> 0$. The the condition in (\ref{E:eqqo}) is valid if and only if $\eta(\theta)\ge\beta$, 
where
\begin{equation}
\eta(\theta)=\psi(\theta)(1-A(\theta)).
\label{E:des}
\end{equation}
It is easy to see that $\eta$ is a strictly increasing function on the 
interval $(0,2\pi)$ (use Lemma \ref{L:mono}). In addition, we have
$$
\lim_{\theta\rightarrow 0}\eta(\theta)=0\quad\mbox{and}\quad\lim_{\theta\rightarrow 2\pi}\eta(\theta)=\infty.
$$

For every $\alpha> 0$, set
\begin{equation}
\eta_{\alpha}(\theta)=\frac{1}{B(\theta)}\left[\frac{A(\theta)^2}{\alpha B(\theta)-1}+1\right]
=\frac{\psi(\theta)^2A(\theta)^2}{\alpha-\psi(\theta)}+\psi(\theta).
\label{E:eta}
\end{equation}
The function $\eta_{\alpha}$ is defined for all $\theta$ such that $0<\theta<\psi^{-1}(\alpha)$. 
It is not hard to see that the function $\eta_{\alpha}$ is strictly increasing on the interval
$(0,\psi^{-1}(\alpha))$ (use Lemma \ref{L:mono}). 

We will next analyze condition (\ref{E:eqqo}) in the case where $\gamma\neq\beta$. First, suppose $\gamma>\beta> 0$. Then condition (\ref{E:eqqo}) can be rewritten
in the following form: $\eta_{\gamma}(\theta)\ge\beta$, or equivalently 
$\theta\ge\eta_{\gamma}^{-1}(\beta)$. Here we assume that $0<\theta<\psi^{-1}(\gamma)$. Note that since
$\eta_{\gamma}(\theta)>\psi(\theta)$, we have $\eta_{\gamma}^{-1}(\beta)<\psi^{-1}(\beta)$.

If $\beta>\gamma> 0$, we can use the fact that condition (\ref{E:eqqo}) does not change if we transpose $\beta$ and $\gamma$, and prove that condition (\ref{E:eqqo}) is equivalent to the condition $\eta_{\beta}(\theta)\ge\gamma$, or equivalently 
$\theta\ge\eta_{\beta}^{-1}(\gamma)$. Here we assume that $0<\theta<\psi^{-1}(\beta)$. Note that since
$\eta_{\beta}(\theta)>\psi(\theta)$, we have $\eta_{\beta}^{-1}(\gamma)<\psi^{-1}(\gamma)$.
\subsection{Main results in the case where $\gamma> 0$. Formulations}\label{SS:mainre}
In this subsection, we formulate a proposition that links the problem of computing the distance to a slanted line $L_{\beta,\gamma}$ with $\gamma> 0$ to minimization problems for the functions $\Lambda^{+}_{\beta,\gamma}$ and
$\Lambda^{+}_{\beta,\gamma}$ given by (\ref{E:anno1}) and (\ref{E:anno2}). 
The structure of the proof of this proposition is as follows. We first characterize the set of those numbers $\theta$, for which the level set $\Gamma_{\theta}$ intersects the given line $L_{\beta,\gamma}$, then select an appropriate number $\Lambda^{+}_{\beta,\gamma}(\theta)$
or $\Lambda^{-}_{\beta,\gamma}(\theta)$ for each admissible value of $\theta$, and next formulate the distance to the line problem as a combination of minimization problems for the functions $\Lambda^{+}_{\beta,\gamma}(\theta)$ and $\Lambda^{-}_{\beta,\gamma}(\theta)$ over the intervals of admissible values of the parameter $\theta$. Note that the resulting formulas depend on the relations between the location parameter $\beta$ and the slope parameter $\gamma$. 

Recall that the functions
$\psi$, $\eta$, and $\eta_{\alpha}$ are defined by (\ref{E:psik}), (\ref{E:des}), and (\ref{E:eta}), respectively.
\begin{theorem}\label{T:the1}
1)\,Let $\beta> 0$. Then
$$
D_{\beta,\beta}=\min\left\{\inf_{\theta\in[\eta^{-1}(\beta),\psi^{-1}(\beta)]}
\left\{\Lambda^{+}_{\beta,\beta}(\theta)\right\},
\inf_{\theta\in[\eta^{-1}(\beta),\psi^{-1}(\beta))}
\left\{\Lambda^{-}_{\beta,\beta}(\theta)\right\}
\right\}.
$$
2)\,Let $\gamma>\beta> 0$. Then
$$
D_{\beta,\gamma}
=\min\left\{\inf_{\theta\in[\eta^{-1}_{\gamma}(\beta),\psi^{-1}(\beta)]}
\left\{\Lambda^{+}_{\beta,\gamma}(\theta)\right\}, 
\inf_{\theta\in[\eta^{-1}_{\gamma}(\beta),\psi^{-1}(\gamma))}
\left\{\Lambda^{-}_{\beta,\gamma}(\theta)\right\}
\right\}.
$$
3)\,Let $\gamma> 0$. Then
$$
D_{0,\gamma} 
=\inf_{\theta\in[0,\psi^{-1}(\gamma))}
\left\{\Lambda^{-}_{0,\gamma}(\theta)\right\}.
$$
4)\,Let $\beta>\gamma> 0$. Then
$$
D_{\beta,\gamma}
=\min\left\{\inf_{\theta\in[\eta_{\beta}^{-1}(\gamma),\psi^{-1}(\beta)]}
\left\{\Lambda^{+}_{\beta,\gamma}(\theta)\right\},
\inf_{\theta\in[\eta_{\beta}^{-1}(\gamma),\psi^{-1}(\gamma))}
\left\{\Lambda^{-}_{\beta,\gamma}(\theta)\right\}
\right\}.
$$
\end{theorem}

It is important to check whether conditions (\ref{E:gg}) and (\ref{E:za}) hold in the minimization problems in the previous theorem. In part 1 of Theorem \ref{T:the1}, we have $\beta=\gamma> 0$ and $\psi(\theta)<\beta\le\eta(\theta)$. Therefore, $A(\theta)^2-(1-\gamma B(\theta))(1-\beta B(\theta))\ge A(\theta)^2-(\eta(\theta)B(\theta)-1)^2\ge 0$. The last inequality follows form the definition of the function $\eta$. In part 2, we use the fact that
condition (\ref{E:za}) is equivalent to the validity of the inequality $\eta_{\gamma}(\theta)\ge\beta$ (see the
remark after formula (\ref{E:eta})). The proof of the validity of condition (\ref{E:za}) in part 4
is similar. Finally, in part 3, we have $\gamma>\beta=0$ and $0\le\psi(\theta)<\gamma$.
This implies that 
$$
A(\theta)^2-(1-\gamma B(\theta))(1-\beta B(\theta))=A(\theta)^2+\frac{\gamma}{\psi(\theta)}-1> 0.
$$

We will next check the validity of condition (\ref{E:gg}) in Theorem \ref{T:the1}. 
If $\theta\le\psi^{-1}(\beta)$, then it is easy to see that (\ref{E:gg}) holds. In the rest of the cases,
we have $\psi^{-1}(\beta)<\theta<\psi^{-1}(\gamma)$ and $v=S^{-}_{\beta,\gamma}(\theta)^2$. Then, we have
$\beta+\gamma v=\beta+\gamma S^{-}_{\beta,\gamma}(\theta)^2$. It follows from $\beta<\psi(\theta)<\gamma$,
$A(\theta)< 0$, $B(\theta)=\frac{1}{\psi(\theta)}$, and (\ref{E:S22}) that
$$
S^{-}_{\beta,\gamma}(\theta)^2\ge\frac{\left(\frac{\gamma}{\psi(\theta)}-1\right)\left(1-\frac{\beta}{\psi(\theta)}
\right)}{\left(\frac{\gamma}{\psi(\theta)}-1\right)^2}=\frac{\psi(\theta)-\beta}
{\gamma-\psi(\theta)}.
$$
Hence, $\beta+\gamma v\ge\beta+\gamma\frac{\psi(\theta)-\beta}
{\gamma-\psi(\theta)}\ge\psi(\theta)$, and condition (\ref{E:gg}) holds.

Theorem \ref{T:the1} will be proved at the end of the next subsection. 
\subsection{Admissible values of $\theta$ and the proof of Theorem \ref{T:the1}}\label{SS:sc}
In the present subsection, we answer the following question: Which level sets $\Gamma_{\theta}$ intersect the given line $L_{\beta,\gamma}$ with $\beta\ge 0$ and $\gamma> 0$. First, several typical situations will be discussed. \\
\\
\it Case 1. \rm Suppose $\gamma=\psi(\theta)$. Then
the condition $(\beta+\gamma v, v)\in\Gamma_{\theta}$ implies that
$\sqrt{v}=S_{\beta}(\theta)$ with $S_{\beta}(\theta)$ given by (\ref{E:S}). 
Since $A(\theta)< 0$ and we should have $\sqrt{v}\ge 0$, the point $(\beta+\gamma v, v)$ in case 1
can belong to the level set $\Gamma_{\theta}$ only if $\beta\ge\psi(\theta)$. Therefore, both formulas
(\ref{E:S2}) and (\ref{E:S}) may be used. \\
\\
\it Case 2. \rm Suppose $0<\gamma<\psi(\theta)$. Then $\sqrt{v}=S^{+}_{\beta,\gamma}(\theta)$ with $S^{+}_{\beta,\gamma}(\theta)$ defined by
(\ref{E:S1}). We choose the solution $S^{+}_{\beta,\gamma}$ because $A(\theta)< 0$, $1-\gamma B(\theta)> 0$, and we should have $\sqrt{v}\ge 0$. We should also assume that
$\psi(\theta)\le\beta$ because otherwise either the expression under the square root sign on the right-hand side of (\ref{E:S1}) is negative, or $\sqrt{v}< 0$. Summarizing what was said above, we see that
if $\gamma<\psi(\theta)$ and $\beta<\psi(\theta)$, then 
$L_{\beta,\gamma}\cap\Gamma_{\theta}=\emptyset$,
while if $\gamma<\psi(\theta)\le\beta$, then the intersection consists of only one point for which
$v=S^{+}_{\beta,\gamma}(\theta)^2$. \\
\\
\it Case 3. \rm Suppose $\gamma>\psi(\theta)$. Then we consider the following three special cases:
\\
\\
\it Case 3a. \rm Let $\gamma>\psi(\theta)$ and $\beta=\psi(\theta)$. Then both solutions
$S^{+}_{\beta,\gamma}$ and $S^{-}_{\beta,\gamma}$ can be used, and the set $L_{\beta,\gamma}\cap\Gamma_{\theta}$ consists of two points 
$(\beta+\gamma v_1,v_1)$ and $(\beta+\gamma v_2,v_2)$ where 
$$
v_1=0\quad\mbox{and}\quad v_2=S^{-}_{\beta,\gamma}(\psi^{-1}(\beta))^2=\frac{4A(\psi^{-1}(\beta))^2}
{(\gamma B(\psi^{-1}(\beta))-1)^2}.
$$
Note that we have
$$
\sqrt{v_1}=0\quad\mbox{and}\quad\sqrt{v_2}=S^{-}_{\beta,\gamma}
(\psi^{-1}(\beta))=\frac{-2A(\psi^{-1}(\beta))}{\gamma B(\psi^{-1}(\beta))-1}.
$$
\\
\it Case 3b. \rm Here $\gamma>\psi(\theta)$ and $\beta<\psi(\theta)$. Taking into account that 
$A(\theta)< 0$, we see that the set $L_{\beta,\gamma}\cap\Gamma_{\theta}$ is not empty only if
the solution $S^{-}_{\beta,\gamma}$ is chosen. Then the intersection $L_{\beta,\gamma}\cap\Gamma_{\theta}$ consists of only one point $(\beta+\gamma v,v)$ for which $v=S^{-}_{\beta,\gamma}(\theta)^2$ (see formula (\ref{E:S22})).
\\
\\
\it Case 3c. \rm Let $\gamma>\psi(\theta)$ and $\beta>\psi(\theta)$. Then the set $L_{\beta,\gamma}\cap\Gamma_{\theta}$ is not empty only when condition (\ref{E:eqqo}) holds,
and both solutions $S^{+}_{\beta,\gamma}(\theta)$ and $S^{-}_{\beta,\gamma}(\theta)$ can be used. Therefore, under the
condition in (\ref{E:eqqo}), the set $L_{\beta,\gamma}\cap\Gamma_{\theta}$ consists of two points
$(\beta+\gamma v_1,v_1)$ and $(\beta+\gamma v_2,v_2)$ for which 
$v_1=S^{+}_{\beta,\gamma}(\theta)^2$ and $v_2=S^{-}_{\beta,\gamma}(\theta)^2$.
Note that if in (\ref{E:eqqo}) we have the equality instead of the inequality, then the two solutions degenerate into one solution
given by
$
\sqrt{v}=\frac{-A(\theta)}{\gamma B(\theta)-1}.
$
Here we also have
$
v=\frac{A(\theta)^2}{(\gamma B(\theta)-1)^2}.
$
\begin{remark} \rm In a sense, condition (\ref{E:eqqo}) controls the distance between the points on $L_{\beta,\gamma}$ and $\Gamma_{\theta}$ which belong to the line $v=0$ (these points are $(\beta,0)$ and 
$(\psi(\theta),0)$, respectively). Condition (\ref{E:eqqo}) also controls the slope $\gamma$ of the line $L_{\beta,\gamma}$ and the slope 
$\psi(\theta)$ of the linear part of the level set $\Gamma_{\theta}$. The slopes mentioned above are taken with respect to the vertical axis $x=0$. 
\end{remark}

Our next goal is to find, for a given slanted line $L_{\beta,\gamma}$, the set $E_{\beta,\gamma}$
of those $\theta$, for which $L_{\beta,\gamma}\cap\Gamma_{\theta}\neq\emptyset$. \\
\\
\underline{\it Lines with $\beta=\gamma> 0$.} \rm Here cases 1 and 3c are relevant. In case 1, $\theta=\psi^{-1}(\beta)$ and $v=0$. In case 3c, $\theta$ is such that $\theta<\psi^{-1}(\beta)$. 
We have to assume here that $\eta(\theta)\ge\beta$, where the function $\eta$ is defined in (\ref{E:des}), since 
for $\beta=\gamma$ the previous inequality is equivalent to condition (\ref{E:eqqo}). It follows that 
$E_{\beta,\beta}=[\eta^{-1}(\beta),\psi^{-1}(\beta)]$. \\
\\
\underline{\it Lines with $\gamma>\beta> 0$.} \rm 
It is not hard to see that in the case where
$\gamma>\beta> 0$, we have to take into account only cases 3a, 3b, and 3c described above. In case 3a, the input into $E_{\beta,\gamma}$ is 
$\theta=\psi^{-1}(\beta)$. In case 3b, the input is $\psi^{-1}(\beta)<\theta<\psi^{-1}(\gamma)$.
Finally in case 3c, we take into account the previous reasoning involving the function $\eta_{\gamma}$.
It follows that the input in this case is $\eta_{\gamma}^{-1}(\beta)\le\theta<\psi^{-1}(\beta)$.
Summarizing what was said above, we obtain the following formula:
$E_{\beta,\gamma}=[\eta_{\gamma}^{-1}(\beta),\psi^{-1}(\gamma))$,
provided that $\gamma>\beta> 0$. \\
\\
\underline{\it Lines with $\gamma>\beta=0$.} \rm In this case, we have
$E_{0,\gamma}=[0,\psi^{-1}(\gamma))$.
For $\theta=0$, the intersection $L_{0,\gamma}\cap\Gamma_0$ is a one-point set $\{(0,0)\}$.
If $0<\theta<\psi^{-1}(\gamma)$, then the set $L_{0,\gamma}\cap\Gamma_0$ also consists of just one point. The corresponding component $v$ and the number $\sqrt{v}$ are given by the same formulas as in case 3b with $\beta=0$.
\\
\\
\underline{\it Lines with $\beta>\gamma> 0$.} \rm For a line $L_{\beta,\gamma}$ with $\beta>\gamma> 0$, 
only case 1, case 2, and case 3c are relevant. In case 1, the input into $E_{\beta,\gamma}$
is $\theta=\psi^{-1}(\gamma)$. In case 2, the input is the interval 
$(\psi^{-1}(\gamma),\psi^{-1}(\beta)]$. Here we take into account Remark \ref{R:drugoj}.
In case 3c, we use the function $\eta_{\beta}$,
and rewrite condition (\ref{E:eqqo}) as follows: 
$\theta\ge\eta_{\beta}^{-1}(\gamma)$. It follows that
$E_{\beta,\gamma}=[\eta_{\beta}^{-1}(\gamma),\psi^{-1}(\beta)]$,
provided that $\beta>\gamma> 0$.
\\
\\
\underline{\it Lines with $\beta> 0$, $\gamma< 0$.} \rm Here only case 2 is relevant. It follows from the geometrical considerations that
$E_{\beta,\gamma}=(-\psi^{-1}(|\gamma|),\psi^{-1}(\beta)]$,
provided that $\beta> 0$ and $\gamma< 0$. It is not hard to see that formulas for $\sqrt{v}$ and $v$ from case 2 hold if $\beta> 0$ and $\gamma< 0$. For $\theta=0$, we have to take $\sqrt{v}=\sqrt{\frac{|\gamma|}{\beta}}$ and $v=\frac{|\gamma|}{\beta}$. The previous equalities are obtained by taking the limit as $\theta\rightarrow 0$ in the formulas for $\sqrt{v}$ and $v$ from case 2.

\it Proof of Theorem \ref{T:the1}. \rm Having collected detailed information on the level sets, intersecting the given line and on the appropriate choice of the functions $S^{+}_{\beta,\gamma}(\theta)$ and $S^{-}_{\beta,\gamma}(\theta)$,
we can finish the proof of Theorem \ref{T:the1}. Indeed, it is not hard to see that Theorem \ref{T:the1} can be derived, using formula (\ref{E:distance}), formula (\ref{E:synge}) with $\sqrt{v}$ equal to $S^{+}_{\beta,\gamma}(\theta)$ or $S^{-}_{\beta,\gamma}(\theta)$ (the choice depends on the relations between $\beta$ and $\gamma$), and taking into account the discussion concerning the admissible values of the parameter $\theta$.
\subsection{Main results in the case where $\gamma< 0$}\label{SS:ls}
In the next statement, we find the distance to a left slanted line.
\begin{theorem}\label{T:gammane}
The following are true: \\
1)\,Suppose $\beta> 0$ and $\gamma< 0$. Suppose also that $\beta>|\gamma|$. Then
$$
D_{\beta,\gamma} 
=\inf_{\theta\in[0,\psi^{-1}(\beta)]}
\left\{\Lambda^{+}_{\beta,\gamma}(\theta)\right\}.
$$
2)\,Suppose $\beta> 0$ and $\gamma< 0$. Suppose also that $\beta<|\gamma|$. Then
$$
D_{\beta,\gamma} 
=\inf_{\theta\in(-\psi^{-1}(|\gamma|),0]}
\left\{\Lambda^{+}_{\beta,\gamma}(\theta)\right\}.
$$
\end{theorem}

\it Proof. \rm Let $\beta> 0$ and $\gamma< 0$. Then, using the same ideas as in the proof of Theorem \ref{T:the1}
and recalling the results on the intersections of the level sets $\Gamma_{\theta}$ with the left slanted line $L_{\beta,\gamma}$ (see Subsection 
\ref{SS:sc}), we establish the following equality:
\begin{equation}
D_{\beta,\gamma} 
=\inf_{\theta\in(-\psi^{-1}(|\gamma|),\psi^{-1}(\beta)]}
\left\{\Lambda^{+}_{\beta,\gamma}(\theta)\right\}.
\label{E:piat}
\end{equation}

The next statement will be used to complete the proof of Theorem \ref{T:gammane}.
\begin{lemma}\label{L:restr}
Let $\beta\ge 0$ and $\gamma< 0$, and suppose $\beta>|\gamma|$. Then the following equality holds:
\begin{align}
&\inf_{\{v:v\ge 0\}}\{d_H((0,1),(\beta+\gamma v,v))\}
=\inf_{\{v:\delta(\beta+\gamma v,v)\ge 0\}}\{d_H((0,1),(\beta+\gamma v,v))\}.
\label{E:restr1}
\end{align}
On the other hand, if $\beta<|\gamma|$, then
\begin{align}
&\inf_{\{v:v\ge 0\}}\{d_H((0,1),(\beta+\gamma v,v))\}
=\inf_{\{v:\delta(\beta+\gamma v,v)\le 0\}}\{d_H((0,1),(\beta+\gamma v,v))\}.
\label{E:restr2}
\end{align}
\end{lemma}
\begin{remark}\label{R:restr} \rm
Lemma \ref{L:restr} states that the minimization problem on the left-hand side of (\ref{E:restr1})
can be reduced to a similar problem on the initial segment of the line $L_{\beta,\gamma}$, connecting the points
$(\beta,0)$ and $\left(0,\frac{\beta}{|\gamma|}\right)$. Similarly, the minimization problem on the left-hand side of (\ref{E:restr2}) reduces to a similar problem on the ray 
$\{(x,v)\in{\cal H}:x=\beta+\gamma v,x\le 0\}$.
Note that the initial point of the ray is the point $\left(0,\frac{\beta}{|\gamma|}\right)$.
\end{remark}

\it Proof of Lemma \ref{L:restr}. \rm If $\beta>|\gamma|$, then the point $\left(0,\frac{\beta}{|\gamma|}\right)$
is above the point $(0,1)$, and the horizontal line $\rho_{\frac{\beta}{|\gamma|}}=\{(x,v):v=\frac{\beta}{|\gamma|}\}$ separates the point
$(0,1)$ from the complement $C$ in $L_{\beta,\gamma}$ of the initial segment mentioned in Remark \ref{R:restr}.
Therefore, 
$$
d_H((0,1);C)\ge d_H\left((0,1);\rho_{\frac{\beta}{|\gamma|}}\right)
=d_H\left((0,1),\left(0,\frac{\beta}{|\gamma|}\right)\right).
$$
We use Lemma \ref{L:horiz} in the proof of the previous equality. Now, it is clear that (\ref{E:restr1}) holds.
The proof of (\ref{E:restr2}) is similar. 

Finally, it is not hard to see that Theorem \ref{T:gammane} follows from (\ref{E:piat}) and Lemma \ref{L:restr}. 
\section{Shorter minimization intervals and simplifications in Theorem \ref{T:the1}}
\label{SS:usef}
The minimization problems appearing in Theorem \ref{T:the1} can be simplified under additional restrictions on the parameters $\beta$ and $\gamma$, or in certain cases, without any additional restrictions. In this section, we employ a combination of several methods. The critical points $P_{\theta(\beta,\gamma)}$ will be used (see the definitions in Lemma \ref{L:curvet} and Remark \ref{R:cu}). We will also compare the functions $\Lambda^{+}_{\beta,\gamma}(\theta)$ and $\Lambda^{-}_{\beta,\gamma}(\theta)$. Moreover, it will be shown that the function $\theta\mapsto\Lambda^{-}_{\beta,\gamma}(\theta)$ increases on a special set. The methods developed in the present section, will allow us to choose shorter minimization intervals in Theorem \ref{T:the1}.
\subsection{Useful facts}\label{SS:st}
Our first goal in the present subsection is to compare the functions $\Lambda^{+}_{\beta,\gamma}$ and $\Lambda^{-}_{\beta,\gamma}$. 
\begin{lemma}\label{L:hoho}
Let $\beta\ge 0$, $\gamma> 0$, $0<\theta< 2\pi$, and suppose the condition 
in (\ref{E:za}) holds. Then
$\Lambda^{+}_{\beta,\gamma}(\theta)\le\Lambda^{-}_{\beta,\gamma}(\theta)$ 
if and only if the following inequality is valid: $2\arctan\gamma\le\theta$.
\end{lemma}

\it Proof. \rm It follows from (\ref{E:anno1}) and (\ref{E:anno2}) that
$$
\Lambda^{+}_{\beta,\gamma}(\theta)-\Lambda^{-}_{\beta,\gamma}(\theta)
=\frac{\theta^2}{1-\cos\theta}\left(S^{+}_{\beta,\gamma}(\theta)-S^{-}_{\beta,\gamma}(\theta)\right)
\left(S^{+}_{\beta,\gamma}(\theta)+S^{-}_{\beta,\gamma}(\theta)-2\cos\frac{\theta}{2}\right).
$$
Next, using (\ref{E:S1}) and (\ref{E:S2}) and simplifying the resulting expressions, we obtain
\begin{align}
\Lambda^{+}_{\beta,\gamma}(\theta)-\Lambda^{-}_{\beta,\gamma}(\theta)&=
\frac{4\theta^2}{1-\cos\theta}\frac{\sqrt{A(\theta)^2-(1-\gamma B(\theta))(1-\beta B(\theta))}}
{(1-\gamma B(\theta))^2} \nonumber \\
&\times\left[A(\theta)-\cos\frac{\theta}{2}(1-\gamma B(\theta))\right].
\label{E:hor1}
\end{align}
Taking into account (\ref{E:gen2}) and (\ref{E:gen3}), we see that
\begin{align}
A(\theta)-\cos\frac{\theta}{2}(1-\gamma B(\theta))&=\frac{\theta\cos\frac{\theta}{2}-2\sin\frac{\theta}{2}}
{\theta-\sin\theta}-\cos\frac{\theta}{2}\left(1-\gamma\frac{1-\cos\theta}{\theta-\sin\theta}\right)
\nonumber \\
&=\frac{2\sin^3\frac{\theta}{2}}{\theta-\sin\theta}\left(\gamma\cot\frac{\theta}{2}-1\right).
\label{E:hor2}
\end{align}
Finally, using (\ref{E:za}), (\ref{E:hor1}), and (\ref{E:hor2}) we see that
$$
\mbox{\rm sign}\left[\Lambda^{+}_{\beta,\gamma}(\theta)-\Lambda^{-}_{\beta,\gamma}(\theta)\right]
=\mbox{\rm sign}\left[\gamma\cot\frac{\theta}{2}-1\right].
$$

Now, it is clear that Lemma \ref{L:hoho} follows from the previous equality.
\begin{corollary}\label{C:derr}
Let $\beta\ge 0$, $\gamma> 0$, $\pi\le\theta< 2\pi$, and suppose the condition in (\ref{E:za}) holds.
Then $\Lambda^{+}_{\beta,\gamma}(\theta)\le\Lambda^{-}_{\beta,\gamma}(\theta)$. 
\end{corollary}
\begin{lemma}\label{L:uses}
For every $\gamma> 0$, $2\arctan\gamma<\min\{\eta^{-1}(\gamma),\pi\}$.
\end{lemma}

\it Proof. \rm The inequality $2\arctan\gamma<\pi$ is trivial, while the inequality 
$2\arctan\gamma<\eta^{-1}(\gamma)$ can be rewritten as follows:
\begin{equation}
\eta(2\alpha)<\tan\alpha,\quad 0<\alpha<\frac{\pi}{2}.
\label{E:equivo}
\end{equation}
Since
$$
\eta(\theta)=\frac{\theta-\sin\theta-\theta\cos\frac{\theta}{2}+2\sin\frac{\theta}{2}}
{1-\cos\theta},
$$
we have
\begin{align*}
&\eta(2\alpha)=\frac{\alpha-\sin\alpha\cos\alpha-\alpha\cos\alpha+\sin\alpha}
{\sin^2\alpha}. \\
\end{align*}
In order to establish the inequality in (\ref{E:equivo}), it suffices to prove that
$$
\frac{\alpha-\sin\alpha\cos\alpha-\alpha\cos\alpha+\sin\alpha}
{\sin^2\alpha}<\frac{\sin\alpha}{\cos\alpha},
$$
or
$$
\alpha\cos\alpha-\alpha\cos^2\alpha+\sin\alpha\cos\alpha<\sin\alpha.
$$
It is easy to see that the previous inequality is equivalent to the well-known inequality
$\alpha<\tan\alpha$, $0<\alpha<\frac{\pi}{2}$.

This completes the proof of Lemma \ref{L:uses}.
\begin{remark}\label{R:iste} \rm
Note that $\eta^{-1}(\gamma)>\pi$ if and only if $\gamma>\frac{\pi}{2}+1$.
\end{remark}
\begin{remark}\label{R:ist} \rm
Since $\eta(\theta)>\psi(\theta)$ (see (\ref{E:des})), we have $\eta^{-1}(\theta)<\psi^{-1}(\theta)$.
Therefore, Lemma \ref{L:uses} implies that for every $\gamma> 0$, $2\arctan\gamma<\psi^{-1}(\gamma)$.
\end{remark}

Recall that $S^{+}_{\beta,\gamma}(\theta)$ and $S^{-}_{\beta,\gamma}(\theta)$ are real numbers only if 
the condition in (\ref{E:za}) holds. Moreover, if $1-\gamma B(\theta)> 0$, then only the number $S^{+}_{\beta,\gamma}(\theta)$ can be positive. On the other hand, if $1-\gamma B(\theta)< 0$, then
both $S^{+}_{\beta,\gamma}(\theta)$ and $S^{-}_{\beta,\gamma}(\theta)$ may be positive. 
In addition, the condition $1-\gamma B(\theta)< 0$ implies the inequality
$S^{+}_{\beta,\gamma}(\theta)\le S^{-}_{\beta,\gamma}(\theta)$.
Finally, if the previous conditions hold and $\theta\ge\pi$, then
$\Lambda^{+}_{\beta,\gamma}(\theta)\le \Lambda^{-}_{\beta,\gamma}(\theta)$ (see Lemma \ref{L:hoho}).
\begin{lemma}\label{L:comparar1}
Suppose $\beta\ge 0$ and $\gamma> 0$. Then the function $\theta\mapsto S^{-}_{\beta,\gamma}(\theta)$ 
is increasing on the set 
$$
E=\left\{\theta\in[0,2\pi):A(\theta)^2-(1-\gamma B(\theta))
(1-\beta B(\theta))\ge 0\right\}\cap\left\{\theta\in[0,2\pi):1-\gamma B(\theta)< 0\right\}.
$$
Moreover, the function $\theta\mapsto\Lambda^{-}_{\beta,\gamma}(\theta)$ is increasing on the set
$[\pi,2\pi)\cap E$.
\end{lemma}

\it Proof. \rm 
It is clear that formula (\ref{E:S2}) implies the following:
$$
S^{-}_{\beta,\gamma}(\theta)=F_{\beta,\gamma}(\theta)+\sqrt{F_{\beta,\gamma}(\theta)^2+G_{\beta,\gamma}
(\theta)},
$$
where
$$
F_{\beta,\gamma}(\theta)=\frac{-A(\theta)}{\gamma B(\theta)-1}\quad\mbox{and}\quad
G_{\beta,\gamma}(\theta)=\frac{1-\beta B(\theta)}{\gamma B(\theta)-1}.
$$
Next, using Lemma \ref{L:mono}, we see that the functions $\theta\mapsto F_{\beta,\gamma}(\theta)$
and $\theta\mapsto G_{\beta,\gamma}(\theta)$ are increasing on the set $E$. Hence, the function $\theta\mapsto S^{-}_{\beta,\gamma}(\theta)$ is increasing on 
the set $E$.

It is clear that the functions 
$\theta\mapsto\frac{\theta^2}{1-\cos\theta}$ and
$\theta\mapsto-\cos\frac{\theta}{2}$
are increasing positive functions on the interval $[\pi,2\pi)$. 
Moreover, we have already established that the function $S^{-}_{\beta,\gamma}$ is increasing on the set
$E$. It follows from (\ref{E:anno2}) that the function $\theta\mapsto
\Lambda^{-}_{\beta,\gamma}(\theta)$ is increasing on the set $[\pi,2\pi)\cap E$.

The proof of Lemma \ref{L:comparar1} is thus completed.

\subsection{Simplifications}\label{SS:simplex} 
Theorem \ref{T:the1} links the distance to a slanted line in the Heston manifold to certain minimization problems for the functions $\Lambda^{+}_{\beta,\gamma}$ and $\Lambda^{-}_{\beta,\gamma}$. In the present subsection, we show that some of the formulas in Theorem \ref{T:the1} can be simplified. Certain additional restrictions may be needed in these simplifications.

Recall that for $0\le\beta<\frac{\pi}{2}$ and $\gamma> 0$, the number $\theta(\beta,\gamma)$ is defined as follows: 
$\theta(\beta,\gamma)=\zeta_{\gamma}^{-1}(\beta)$, where the function $\zeta_{\gamma}$ is given by
(\ref{E:solvab1}). Moreover, for $\frac{\pi}{2}\le\beta$ and $\gamma> 0$, we put $\theta(\beta,\gamma)=\psi^{-1}
(\beta)$ (see Lemma \ref{L:curvet} and Remark \ref{R:cu}).

Our first result in the present subsection simplifies part 1 of Theorem \ref{T:the1}.
\begin{corollary}\label{C:11}
Let $\beta> 0$. Then
$$
D_{\beta,\beta}=\inf_{\theta\in[\eta^{-1}(\beta),\psi^{-1}(\beta)]}
\left\{\Lambda^{+}_{\beta,\beta}(\theta)\right\}.
$$
\end{corollary}

\it Proof. \rm The inequality $2\arctan\beta<\eta^{-1}(\beta)$ follows from Lemma \ref{L:uses}. Moreover, the values of the parameter $\theta$ in the minimization problems in part 1 of Theorem \ref{T:the1} satisfy the condition $\eta^{-1}(\beta)\le\theta$. Now, Lemma \ref{L:hoho} implies that for such values of $\theta$, $\Lambda^{+}_{\beta,\beta}(\theta)\le\Lambda^{-}_{\beta,\beta}(\theta)$. Therefore,  Corollary \ref{C:11} hods.

For $0<\beta<\frac{\pi}{2}$, we can shorten the minimization interval in Corollary \ref{C:11}.
\begin{corollary}\label{C:1}
Let $0<\beta<\frac{\pi}{2}$. Then
$$
D_{\beta,\beta}=\inf_{\theta\in[\eta^{-1}(\beta),\theta(\beta,\beta)]}
\left\{\Lambda^{+}_{\beta,\beta}(\theta)\right\}.
$$
\end{corollary}

\it Proof. \rm The proof of Corollary \ref{C:1} is a mixture of analytical and geometrical methods. Consider the line $L_{\beta,\beta}$ with $0<\beta<\frac{\pi}{2}$. A geometrical analysis of the result in part 1 of Theorem \ref{T:the1} leads us to the conclusion that if $\theta=\eta^{-1}(\beta)$, then the line $L_{\beta,\beta}$ is tangent to the level set $\Gamma_{\theta}$. In this case, we have $S^{+}_{\beta,\beta}(\theta)=S^{-}_{\beta,\beta}(\theta)$. 
On the other hand, for every $\theta$ with $\eta^{-1}(\beta)<\theta<\psi^{-1}(\beta)$, the level set $\Gamma_{\theta}$
intersects the line $L_{\beta,\beta}$ twice, and the number $S^{+}_{\beta,\beta}(\theta)$ corresponds to the intersection point that has the smaller value of the component $v$. Finally, if $\theta=\psi^{-1}(\beta)$, then 
$S^{+}_{\beta,\beta}(\theta)=S_{\beta}(\theta)$ is chosen, where $S_{\beta}$ is defined in (\ref{E:S}).

The line $L_{\beta,\beta}$ intersects the curve described in (\ref{E:curve1}) at exactly one point $P_{\theta(\beta,\beta)}$ (see Lemma \ref{L:curvet}). Put $\theta_0=\theta(\beta,\beta)$ and consider the 
level set $\Gamma_{\theta_0}$. The line $L_{\beta,\beta}$ is a secant of the convex set $\Gamma_{\theta_0}$.
Denote by $L^{-}_{\beta,\beta}$ the part of $L_{\beta,\beta}$ that is separated from the point $(0,1)$ 
by the level set $\Gamma_{\theta_0}$, and by $L^{+}_{\beta,\beta}$ the remaining part of $L_{\beta,\beta}$.
It follows from the definition of the Riemannan distance that
\begin{equation}
d_H((0,1),L^{-}_{\beta,\beta})\ge d_H\left((0,1),\Gamma_{\theta_0}\right).
\label{E:price1}
\end{equation}
Therefore,
$
d_H((0,1),L^{-}_{\beta,\beta})\ge d_H\left((0,1),P_{\theta_0}\right)
$
(the previous inequality follows from (\ref{E:011}) and (\ref{E:price1})). Since $P_{\theta_0}
\in L^{-}_{\beta,\beta}$, we obtain $d_H((0,1),L^{-}_{\beta,\beta})=d_H\left((0,1),P_{\theta_0}\right)$.
Hence,
\begin{equation}
d_H((0,1),L_{\beta,\beta})=d_H((0,1),L^{+}_{\beta,\beta}).
\label{E:price2}
\end{equation}
It follows from (\ref{E:price2}) that the minimization intervals in part 1 of Theorem \ref{T:the1} can be reduced to the interval $[\eta^{-1}(\beta),\theta(\beta,\beta)]$. Finally, we can remove the function $\Lambda^{-}_{\beta,\beta}$
from the resulting formula, reasoning as in the proof of Lemma \ref{L:1}.

This completes the proof of Corollary \ref{C:1}.

The next corollary states that we can choose a longer, but simpler, minimization interval in Corollary \ref{C:1}.
\begin{corollary}\label{C:111}
Let $0<\beta<\frac{\pi}{2}$. Then
$$
D_{\beta,\beta}=\inf_{\theta\in[\eta^{-1}(\beta),2\beta]}
\left\{\Lambda^{+}_{\beta,\beta}(\theta)\right\}.
$$
\end{corollary}

\it Proof. \rm Taking into account Corollaries \ref{C:1} and \ref{C:111}, we see that it suffices to prove the following estimates:
\begin{equation}
\theta(\beta,\beta)< 2\beta<\psi^{-1}(\beta),\quad 0<\beta<\frac{\pi}{2}.
\label{E:itd}
\end{equation}
Since $\theta(\beta,\beta)=\zeta^{-1}_{\beta}(\beta)$, the estimates in (\ref{E:itd}) can be rewritten in the following equivalent form:
\begin{equation}
\psi(2\beta)<\beta<\zeta_{\beta}(2\beta),\quad 0<\beta<\frac{\pi}{2}.
\label{E:itd1}
\end{equation}

It is known that 
\begin{equation}
\sin\beta>\beta\cos\beta,\quad\mbox{for all}\quad 0<\beta<\frac{\pi}{2}.
\label{E:know}
\end{equation} 
Next, using (\ref{E:solvab1}) and (\ref{E:know}), we obtain
\begin{align*}
\zeta_{\beta}(2\beta)&=\frac{2\beta+\sin(2\beta)-\beta(1+\cos(2\beta))}{2}
=\beta+\sin\beta\,\cos\beta-\beta\cos^2\beta \\
&=\beta\sin^2\beta+\sin\beta\cos\beta>\beta.
\end{align*}
This establishes the second inequality in (\ref{E:itd1}).

To establish the first inequality in (\ref{E:itd1}), we use (\ref{E:psik}) and (\ref{E:know}). This gives
\begin{align*}
\psi(2\beta)=\frac{2\beta-\sin(2\beta)}{1-\cos(2\beta)}=\frac{\beta-\sin\beta\,\cos\beta}{\sin^2\beta}<\beta.
\end{align*}

The proof of Corollary \ref{C:111} is thus completed.

The next statement simplifies part 2 of Theorem \ref{T:the1}.
\begin{corollary}\label{C:2}
Let $\gamma>\beta> 0$. Then
$$
D_{\beta,\gamma}
=\min\left\{\inf_{\theta\in[\eta^{-1}_{\gamma}(\beta),\theta(\beta,\gamma)]}
\left\{\Lambda^{+}_{\beta,\gamma}(\theta)\right\}, 
\inf_{\theta\in[\eta^{-1}_{\gamma}(\beta),\theta(\beta,\gamma)]}
\left\{\Lambda^{-}_{\beta,\gamma}(\theta)\right\}
\right\}.
$$
\end{corollary}

\it Proof. \rm The proof of Corollary \ref{C:2} is similar to that of Corollary \ref{C:1}. 
Let $\theta=\eta_{\gamma}^{-1}(\beta)$. Then the line $L_{\beta,\gamma}$ is tangent to the level set $\Gamma_{\theta}$. In this case, we have $S^{+}_{\beta,\gamma}(\theta)=S^{-}_{\beta,\gamma}(\theta)$. 
On the other hand, for every $\theta$ with $\eta^{-1}_{\gamma}(\beta)<\theta\le\psi^{-1}(\beta)$, the level set $\Gamma_{\theta}$
intersects the line $L_{\beta,\gamma}$ twice, and the number $S^{+}_{\beta,\gamma}(\theta)$ corresponds to the intersection point that has the smaller value of the component $v$. Finally, if $\psi^{-1}(\beta)<\theta<\psi^{-1}(\gamma)$, then the level set $\Gamma_{\theta}$
intersects the line $L_{\beta,\gamma}$ once, and the number
$S^{-}_{\beta,\gamma}(\theta)$ has to be chosen. The line $L_{\beta,\gamma}$ contains one and only one critical point $P_{\theta(\beta,\gamma)}$. For $0<\beta<\frac{\pi}{2}$, this follows from Lemma \ref{L:curvet}, while for $\frac{\pi}{2}\le\beta$, we have $\theta(\beta,\gamma)=\psi^{-1}(\beta)$. 

Put $\theta_0=\theta(\beta,\gamma)$, and consider the 
level set $\Gamma_{\theta_0}$. The line $L_{\beta,\gamma}$ is a secant of the convex set $\Gamma_{\theta_0}$.
Denote by $L^{-}_{\beta,\gamma}$ the part the line of $L_{\beta,\gamma}$ that is separated from the point $(0,1)$ 
by the level set $\Gamma_{\theta_0}$, and by $L^{+}_{\beta,\gamma}$ the remaining part of $L_{\beta,\gamma}$.
It follows from the definition of the Riemannan distance that
\begin{equation}
d_H((0,1),L^{-}_{\beta,\gamma})\ge d_H\left((0,1),\Gamma_{\theta_0}\right).
\label{E:price10}
\end{equation}
Therefore,
$
d_H((0,1),L^{-}_{\beta,\gamma})\ge d_H\left((0,1),P_{\theta_0}\right)
$
(the previous inequality follows from (\ref{E:011}) and (\ref{E:price10})). Since $P_{\theta_0}
\in L^{-}_{\beta,\gamma}$, we obtain $d_H((0,1),L^{-}_{\beta,\gamma})=d_H\left((0,1),P_{\theta_0}\right)$.
Hence,
\begin{equation}
d_H((0,1),L_{\beta,\gamma})=d_H((0,1),L^{+}_{\beta,\gamma}).
\label{E:price20}
\end{equation}
It follows from (\ref{E:price2}) that the minimization intervals in part 2 of Theorem \ref{T:the1} can be reduced to the interval $[\eta^{-1}_{\gamma}(\beta),\theta(\beta,\gamma)]$. 

This completes the proof of Corollary \ref{C:2}.

The next assertion shows that under additional restrictions on the parameters $\beta$ and $\gamma$, the function
$\Lambda^{-}_{\beta,\gamma}$ can be removed from the formula in Corollary \ref{C:2}.
\begin{corollary}\label{C:firm1}
Suppose $\beta>\frac{\pi}{2}$ and $\gamma>\frac{\pi}{2}+\frac{2}{2\beta-\pi}$. Then
$$
D_{\beta,\gamma}
=\inf_{\theta\in[\eta^{-1}_{\gamma}(\beta),\psi^{-1}(\beta)]}
\left\{\Lambda^{+}_{\beta,\gamma}(\theta)\right\}.
$$
\end{corollary}

\it Proof. \rm It is not hard to see that if $\beta>\frac{\pi}{2}$, then
$\frac{\pi}{2}+\frac{2}{2\beta-\pi}>\beta$.
Therefore, we have $\gamma>\beta$.
Moreover, it is clear that the conditions in Corollary \ref{C:firm1} are equivalent to the following inequalities: 
$\gamma>\beta>\frac{\pi}{2}$ and $\pi<\eta^{-1}_{\gamma}(\beta)$. Next, using the formula in part 2 of Theorem \ref{T:the1} and the second statement in Lemma \ref{L:comparar1}, we obtain
$$
D_{\beta,\gamma}=\min\left\{\inf_{\theta\in[\eta^{-1}_{\gamma}(\beta),\psi^{-1}(\beta)]}
\left\{\Lambda^{+}_{\beta,\gamma}(\theta)\right\}, 
\Lambda^{-}_{\beta,\gamma}(\eta^{-1}_{\gamma}(\beta))\right\}.
$$
Moreover, since $\pi<\eta^{-1}_{\gamma}(\beta)$, Corollary \ref{C:derr} implies that 
$\Lambda^{+}_{\beta,\gamma}(\eta^{-1}_{\gamma}(\beta))\le\Lambda^{-}_{\beta,\gamma}(\eta^{-1}_{\gamma}(\beta))$.

The proof of Corollary \ref{C:firm1} is thus completed.

In the next assertion, we simplify the formula in part 3 of Theorem \ref{T:the1}.
\begin{corollary}\label{C:firm2}
(a)\,Let $0<\gamma<\frac{\pi}{2}$. Then
$$
D_{0,\gamma} 
=\inf_{\theta\in[0,2\gamma]}
\left\{\Lambda^{-}_{0,\gamma}(\theta)\right\}.
$$ 
(b)\,Let $\gamma\ge\frac{\pi}{2}$. Then
$$
D_{0,\gamma} 
=\inf_{\theta\in[0,\pi]}
\left\{\Lambda^{-}_{0,\gamma}(\theta)\right\}.
$$
\end{corollary}

\it Proof. \rm Let $0<\gamma<\frac{\pi}{2}$, and suppose we travel along the line $L_{0,\gamma}$ from the point
$(0,0)$. Then it is not hard to see that the values of the parameters $\theta$ and $v$ increase. By Lemma \ref{L:incr}, the part of the line $L_{0,\gamma}$, lying above the horizontal line $y=1$, can be disregarded in the minimization problem in the formula in part 3 of Theorem \ref{T:the1}. Therefore,
\begin{equation}
D_{0,\gamma} 
=\inf_{\theta\in[0,\theta_0]}
\left\{\Lambda^{-}_{0,\gamma}(\theta)\right\},
\label{E:ekr}
\end{equation}
where $\theta_0=\delta((0,1),(\gamma,1))$. Our next goal is to prove that
\begin{equation}
\theta_0\le 2\gamma\le\psi^{-1}(\gamma).
\label{E:hr1}
\end{equation}
It follows from (\ref{E:nes6}) that $\theta_0=f^{-1}(\gamma)$, where 
\begin{align*}
&f(\delta)=\frac{\delta-\delta\cos\left(\frac{\delta}{2}\right)+2\sin\left(\frac{\delta}{2}\right)-\sin(\delta)}
{\sin^2\left(\frac{\delta}{2}\right)}.
\end{align*}

For all $0<\delta<2\pi$, we have
\begin{align*}
&f(\delta)
=\frac{\left(1-\cos\left(\frac{\delta}{2}\right)\right)\left(\delta+2\sin\left(\frac{\delta}{2}\right)\right)}
{\sin^2\left(\frac{\delta}{2}\right)}
=\frac{\delta+2\sin\left(\frac{\delta}{2}\right)}{1+\cos\left(\frac{\delta}{2}\right)}\ge\frac{\delta}{2}.
\end{align*}
Therefore, $f^{-1}(\tau)\le 2\tau$ for all $0<\tau<\frac{\pi}{2}$, and hence $\theta_0\le 2\gamma$. This establishes the first inequaltiy in (\ref{E:hr1}). The second inequality in (\ref{E:hr1}) follows from (\ref{E:xx}). Now, it is clear that part $(a)$ of Corollary \ref{C:firm2} follows from (\ref{E:hr1}), (\ref{E:ekr}), 
and the formula in part 3 of Theorem \ref{T:the1}.

We will next prove part (b) of Corollary \ref{C:firm2}. First, we observe that the inequalities $\gamma\ge\frac{\pi}{2}$ and $\psi^{-1}(\gamma)\ge\pi$ are equivalent
(see Remark \ref{R:iste}). It is not hard to see that part (b) of Corollary \ref{C:firm2}
follows from part 3 of Theorem \ref{T:the1} and the second statement in Lemma \ref{L:comparar1}.
\section{Distance to special lines in the Heston manifold}\label{S:dsl}
There are special lines in the Heston manifold, for which the distance formulas are extremely simple. Fix $0<\theta<\pi$, and consider the level set $\Gamma_{\theta}$. Recall that we denoted by $P_{\theta}$ 
the critical point located on $\Gamma_{\theta}$ (see (\ref{E:curve1})). Then, the special line $T_{\theta}$ is defined as the tangent line to $\Gamma_{\theta}$ at the point $P_{\theta}$. In the next theorem, we compute the distance from the point $(0,1)$ to the line $T_{\theta}$, and also find the numbers $\beta(\theta)$ and $\gamma(\theta)$ such that 
$T_{\theta}=L_{\beta(\theta),\gamma(\theta)}$.
\begin{theorem}\label{T:di} 
1)\,Let $0<\theta<\pi$. Then $d_H((0,1),T_{\theta})=\theta$. \\
\\
2)\, The line $T_{\theta}$ coincides with the line $L_{\beta(\theta),\gamma(\theta)}$ where
$\beta(\theta)=\frac{\theta}{2}$ and $\gamma(\theta)=\tan\frac{\theta}{2}$.
\end{theorem}
  
\it Proof. \rm The level curve $\Gamma_{\theta}$ is convex, and hence it separates the point $(0,1)$ from the tangent
line $T_{\theta}$. It follows that the distance from $(0,1)$ to $T_{\theta}$ exceeds the distance from
$(0,1)$ to $\Gamma_{\theta}$. Moreover, the latter distance is equal to $\theta$ (see the first formula in Theorem \ref{T:ditol}), and it is attained at the point $P_{\theta}$, belonging to the line $T_{\theta}$. Now, it is clear that the previous reasoning implies the equality in part 1 of Theorem \ref{T:di}. 

We will next prove part 2 of Theorem \ref{T:di}. The tangent line $T_{\theta}$ passes through the point
$$
P_{\theta}=\left(\frac{\theta+\sin\theta}{2},\cos^2\frac{\theta}{2}\right)
$$ 
(for the previous formula, see (\ref{E:curve1})). The slope $\tau$ of the tangent line at $P_{\theta}$ is given by
$$
\tau=\frac{\partial v}{\partial x}\left(\frac{\theta+\sin\theta}{2}\right),
$$
where $x\mapsto v(x)$ is the function describing the level set $\Gamma_{\theta}$. Recall that
$$
\frac{\partial v}{\partial x}=\frac{1-\cos\theta}{\theta-\sin\theta}\left\{1+
\frac{\theta\cos\frac{\theta}{2}-2\sin\frac{\theta}{2}}{\sqrt{N(\theta,x)}}\right\}
$$
for all $\psi(\theta)\le x<\infty$ (see (\ref{E:rirr})). In the previous formula, 
\begin{align}
&N(\theta,x)=\sin^2\frac{\theta}{2}\left[2(\theta-\sin\theta)x+2(1-\cos\theta)-\theta^2\right]
\label{E:soloto}
\end{align} 
(see (\ref{E:solot})). Plugging $x=\frac{\theta+\sin\theta}{2}$ into formula (\ref{E:soloto}) and making simplifications, we obtain
$$
\sqrt{N\left(\theta,\frac{\theta+\sin\theta}{2}\right)}=2\sin^3\frac{\theta}{2}.
$$
Therefore
\begin{align*}
\tau=\frac{2\sin^2\frac{\theta}{2}}{\theta-\sin\theta}\left[\frac{2\sin^3\frac{\theta}{2}+\theta\cos\frac{\theta}{2}-2\sin\frac{\theta}{2}}{2\sin^3\frac{\theta}{2}}\right]=\frac{1}{\sin\frac{\theta}{2}(\theta-\sin\theta)}
\left[\theta\cos\frac{\theta}{2}-2\sin\frac{\theta}{2}\cos^2\frac{\theta}{2}\right]=\cot\frac{\theta}{2}.
\end{align*}
It follows that the equation of the tangent line $T_{\theta}$ is
$
v-\cos^2\frac{\theta}{2}=\cot\frac{\theta}{2}\left(x-\frac{\theta+\sin\theta}{2}\right),
$
or equivalently,
$
x=v\tan\frac{\theta}{2}+\frac{\theta}{2}.
$

This completes the proof of Theorem \ref{T:di}.
\begin{remark}\label{R:final} \rm
Let $\pi\le\theta< 2\pi$. Then the minimum distance from the point
$(0,1)$ to the level set $\Gamma_{\theta}$ is equal to $\frac{\theta}{\sin\frac{\theta}{2}}$, and the minimum is attained at the point $(\psi(\theta),0)$ where $\psi(\theta)
=\frac{\theta-\sin\theta}{1-\cos\theta}$ (see Lemma \ref{L:critic} and Theorem \ref{T:ditol}). The one-sided tangent line to $\Gamma_{\theta}$ at the point $(\psi(\theta),0)$ is horizontal (see Remark \ref{R:tan}). Next, reasoning as in the proof of part 1 of Theorem \ref{T:di}, we obtain the following equality: 
$$
d_H((0,1),[\psi(\theta),\infty))=\frac{\theta}{\sin\frac{\theta}{2}}.
$$ 
Note that the previous equality also follows from formula (\ref{E:012}) and Lemma \ref{L:horiz}.
\end{remark}
\section{Distance to the line and the small-time limit of the implied volatility}\label{S:appii}
Let $C$ be the call pricing function in the Heston model. The implied volatility $I$ associated with $C$ is determined from the following formula:
\begin{equation}
C^{BS}(T,K,I(T,K,x_0,v_0,\rho))=C(T,K,x_0,v_0,\rho).
\label{E:iv}
\end{equation}
In (\ref{E:iv}), $T> 0$ is the maturity of the call option $C$ and $K> 0$ is the strike. The symbol on the left-hand side of (\ref{E:iv}) stands for the call pricing function in the Black-Scholes model evaluated at $T$, $K$, and the volatility parameter equal to $I(T,K,x_0,v_0,\rho)$. We refer the reader to \cite{G} for more information on call pricing functions and the implied volatility. Set
$$
I_H(K,x_0,v_0,\rho)=\lim_{T\rightarrow 0}I(T,K,x_0,v_0,\rho).
$$
The function $I_H$ is the leading term in the asymptotic expansion of the implied volatility at small maturities. 

It is known that under certain restrictions, the following formula holds for $K\neq S_0$:
\begin{equation}
I_H(K,x_0,v_0,\rho)=\frac{\left|\log\frac{x_0}{K}\right|}
{\inf_{v_1\ge 0}\left\{d_H^{(\rho)}\left(\left(\log\frac{x_0}{K},v_0\right),(0,v_1)\right)\right\}}.
\label{E:term1}
\end{equation}
\begin{remark}\label{R:rem1} \rm It is not hard to see, using (\ref{E:sveli}) and (\ref{E:form2}), that 
$$
d_H^{(\rho)}\left(\left(\log\frac{x_0}{K},v_0\right),(0,v_1)\right)=d_H^{(\rho)}\left(\left(0,v_0\right),
(\log\frac{K}{x_0},v_1)\right).
$$
Therefore, the expression in the denominator of the fraction on the right-hand side of (\ref{E:term1}) can be replaced
by the expression $\inf_{v_1\ge 0}\left\{d_H^{(\rho)}\left(\left(\log\frac{K}{x_0},v_0\right),(0,v_1)\right)\right\}$.
\end{remark}
\begin{remark}\label{R:rem2} \rm
It is assumed in (\ref{E:term1}) that $v_0\neq 0$. Formula (\ref{E:term1}) links the small maturity behavior of the implied volatility in the correlated Heston model with the distance from a point to a vertical line in the Riemannian manifold associated with the model. Note that formula (\ref{E:term1}) is meaningful only under the condition $K\neq S_0$, which corresponds to out-of-the-money or in-the-money options. The case of at-the-money options is qualitatively different.
\end{remark}

Formula (\ref{E:term1}) and more general formulas, from which (\ref{E:term1}) follows, can be found in various publications. We only mention the book \cite{HL} by P. Henry-Labord\`{e}re (see formula 6.25 in this book), the paper \cite{BBF} of H. Berestycki, J. Busca, and I. Florent (see Subsection 6.2 in \cite{BBF} for the results concerning the Heston model), and a lecture \cite{L} of A. L. Lewis. In the paper \cite{FJ} of M. Forde and A. Jacquier, the small maturity limit of the implied volatility is represented in terms of the Legendre-Fenchel transform of the limiting cumulant generating function of the Heston density (see \cite{FJ}, Theorem 2.4; see also \cite{FJL}). Note that in \cite{FJ}, the non-hitting condition $c^2< 2a$ for the variance process is assumed, while in \cite{FJL}, another restriction on the parameters is imposed. In a recent pre-print \cite{AFLZ}, the small maturity asymptotics of the implied volatility are studied in general uncorrelated local-stochastic volatility models. 
For the uncorrelated Heston model, formula (\ref{E:term1}) can be obtained from formula (20) in \cite{AFLZ}.
We would also like to mention the papers \cite{DFJV1,DFJV2,FFF,JR} dealing with similar and related problems for the Heston model and some other stochastic volatility models.

It will be shown in the next standard lemma that the problem of computing the small maturity limit of the implied volatility in the Heston model can be reduced to that of computing the distance from the point $(0,1)$ to a vertical or slanted line in the corresponding uncorrelated model.
\begin{lemma}\label{L:reduce}
The following formula holds:
\begin{equation}
I_H(K,x_0,v_0,\rho)=\frac{c\left|\log\frac{x_0}{K}\right|}{\sqrt{v_0}\widehat{D}_{\beta,\gamma}},
\label{E:inad}
\end{equation}
where
$$
\beta=\frac{c\log\frac{K}{S_0}}{v_0\sqrt{1-\rho^2}}
+\frac{\rho}{\sqrt{1-\rho^2}}\quad\mbox{and}\quad\gamma=-\frac{\rho}{\sqrt{1-\rho^2}}.
$$
\end{lemma}

\it Proof. \rm We will prove that
\begin{equation}
\inf_{v_1\ge 0}\left\{d_H^{(\rho)}\left(\left(\log\frac{x_0}{K},v_0\right),(0,v_1)\right)\right\}
=\frac{\sqrt{v_0}}{c}\widehat{D}_{\beta,\gamma}. 
\label{E:ro1}
\end{equation}
Using (\ref{E:sveli}), we get
\begin{equation}
d_H^{(\rho)}\left(\left(\log\frac{x_0}{K},v_0\right),(0,v_1)\right)
=\frac{1}{c}d_H\left(\left(\frac{c\log\frac{x_0}{K}-\rho v_0}{\sqrt{1-\rho^2}},v_0\right),\left(\frac{-\rho v_1}{\sqrt{1-\rho^2}},v_1\right)\right).
\label{E:app}
\end{equation}
Next, taking into account (\ref{E:form2}), (\ref{E:form3}), and (\ref{E:app}), we obtain
\begin{align}
&d_H^{(\rho)}\left(\left(\log\frac{x_0}{K},v_0\right),(0,v_1)\right)
=\frac{\sqrt{v_0}}{c}d_H\left((0,1),\left(\frac{c\log\frac{K}{x_0}-\rho(v_1-v_0)}
{v_0\sqrt{1-\rho^2}},\frac{v_1}{v_0}\right)\right).
\label{E:svelico}
\end{align}
It follows from (\ref{E:svelico}) that 
\begin{align*}
&\inf_{v_1\ge 0}\left\{d_H^{(\rho)}\left(\left(\log\frac{x_0}{K},v_0\right),(0,v_1)\right)\right\}
=\frac{\sqrt{v_0}}{c}\inf_{v_1\ge 0}\left\{d_H\left((0,1),\left(\frac{c\log\frac{K}{x_0}-\rho(v_1-v_0)}
{v_0\sqrt{1-\rho^2}},\frac{v_1}{v_0}\right)\right)\right\} \\
&=\frac{\sqrt{v_0}}{c}\inf_{v\ge 0}\left\{d_H\left((0,1),\left(\frac{c\log\frac{K}{x_0}-\rho(v_0v-v_0)}
{v_0\sqrt{1-\rho^2}},v\right)\right)\right\}=\frac{\sqrt{v_0}}{c}\widehat{D}_{\beta,\gamma},
\end{align*}
where $\beta$ and $\gamma$ are defined above. This establishes formula (\ref{E:ro1}). Finally, formula (\ref{E:inad}) can be obtained from (\ref{E:term1})
and (\ref{E:ro1}).

The proof of Lemma \ref{L:reduce} is thus completed.

\end{document}